\documentclass[pra,twocolumn,showpacs,preprintnumbers,amsmath,amssymb]{revtex4}
\usepackage{graphicx}
\usepackage{dcolumn}
\usepackage{bm}
\newcommand{\p}{\partial}
\newcommand{\vek}[1]{\bm{\mathrm{#1}}}
\newcommand{\eq}{\mathit{eq}}
\newcommand{\pot}{\mathit{pot}}
\newcommand{\kin}{\mathit{kin}}
\newcommand{\cl}{\mathit{cl}}
\newcommand{\hd}{\mathit{hd}}
\newcommand{\inter}{\mathit{int}}
\newcommand{\Eq}[1]{Eq.\@ (\ref{#1})}
\newcommand{\Eqs}[1]{Eqs.\@ (\ref{#1})}
\newcommand{\Ref}[1]{Ref.\@ \cite{#1}}
\newcommand{\Refs}[1]{Refs.\@ \cite{#1}}
\newcommand{\Fig}[1]{Fig.\@ \ref{#1}}
\newcommand{\Figs}[1]{Figs.\@ \ref{#1}}
\newcommand{\Sec}[1]{Sec.\@ \ref{#1}}
\newcommand{\Tab}[1]{Table\@ \ref{#1}}
\DeclareMathOperator{\Real}{Re}
\DeclareMathOperator{\Imag}{Im}
\renewcommand{\Re}{\Real}
\renewcommand{\Im}{\Imag}
\begin{document}


\title{Collective modes of trapped Fermi gases with in-medium interaction}

\author{S.Chiacchiera}
\altaffiliation[Present address: ]{Centro de F{\'i}sica Computacional, Department of Physics,
University of Coimbra, P-3004-516 Coimbra, Portugal}
\affiliation{Universit{\'e} de Lyon, F-69622 Lyon, France;
  Univ. Lyon 1, Villeurbanne;
  CNRS/IN2P3, UMR5822, IPNL}
\author{T.Lepers}
\email{t.lepers@ipnl.in2p3.fr}
\affiliation{Universit{\'e} de Lyon, F-69622 Lyon, France;
  Univ. Lyon 1, Villeurbanne;
  CNRS/IN2P3, UMR5822, IPNL}
\author{D.Davesne}
\affiliation{Universit{\'e} de Lyon, F-69622 Lyon, France;
  Univ. Lyon 1, Villeurbanne;
  CNRS/IN2P3, UMR5822, IPNL}
\author{M.Urban}
\affiliation{Institut de Physique Nucl{\'e}aire, CNRS and
  Universit\'e Paris-Sud 11, 91406 Orsay Cedex, France}%

\date{December 1, 2008}
\begin{abstract}
Due to Pauli blocking of intermediate states, the scattering matrix
(or $T$ matrix) of two fermionic atoms in a Fermi gas becomes
different from that of two atoms in free space. This effect becomes
particularly important near a Feshbach resonance, where the
interaction in free space is very strong but becomes effectively
suppressed in the medium. We calculate the in-medium $T$ matrix in
ladder approximation and study its effects on the properties of
collective modes of a trapped gas in the normal-fluid phase. We introduce the
in-medium interaction on both sides of the Boltzmann equation, namely
in the calculation of the mean field and in the calculation of the
collision rate. This allows us to explain the observed upward shift of the frequency of the quadrupole
mode in the collisionless regime. By including the mean field, we also improve
considerably the agreement with the measured temperature dependence of frequency and
damping rate of the scissors mode, whereas the use of the in-medium cross section
deteriorates the description, in agreement with previous work.
\end{abstract}

\maketitle

\section{\label{sec:intro}Introduction}

In the last few years, experiments on collective modes in ultracold
trapped Fermi gases at Duke University \cite{KinastHemmer,
KinastTurlapov} and at Innsbruck \cite{BartensteinAltmeyer, Altmeyer,Altmeyer2,
Wright, RiedlBruun} revealed a lot of interesting information about
the different regimes which can be realized in these systems. This is
mainly due to the possibility to change not only the temperature of
the gas, but also the interaction between the atoms by adjusting the
magnetic field in the vicinity of a Feshbach resonance. One of the
objectives of these experiments was to detect the transition from the
normal to the superfluid phase. However, although the frequencies of
the collective modes are generally different in the hydrodynamic and
in the collisionless regime, it was recognized that a hydrodynamic
behaviour of the gas is not an unambiguous sign of superfluidity. In
fact, hydrodynamic behavior can be a consequence of superfluidity, but
also of a sufficiently high collision rate in the normal-fluid phase
\cite{Cozzini}. Recently it was therefore proposed to distinguish the
superfluid, the collisionally hydrodynamic, and the collisionless
regime \cite{Wright}.

The experimental results suggest that the transition from the
superfluid to any of the normal-fluid regimes is always accompanied by
a very strong damping of the collective modes. While the frequencies
of the collective modes in the zero-temperature limit can easily be
predicted in the framework of superfluid hydrodynamics, literature on
the finite-temperature case is relatively sparse. One suggestion put
forward by Griffin et al. was to apply Landau's two-fluid
hydrodynamics \cite{TaylorGriffin}, but this requires that the
collision rate in the normal-fluid component is sufficiently high,
i.e., much higher than the mode frequency. This is usually not the
case, since in a trapped system the collective modes have frequencies
which are at least of the order of the trap frequency. In the opposite
limit, i.e., if the normal-fluid component is collisionless, as it
should be in the weakly interacting limit, a quasiparticle transport
theory which couples superfluid hydrodynamics to a Vlasov equation for
the normal component was suggested by one of the authors
\cite{UrbanSchuck, Urban2007, Urban2008}. Although this theory
provides a qualitative explanation of the damping of the collective
modes near the transition from the superfluid to the collisionless
normal-fluid phase, it cannot give a quantitative description of the
recent experiments since these are generally done in the strongly
interacting regime (near a Feshbach resonance).

In the present paper, we will consider the case of attractive interaction, i.e.,
negative scattering length $a < 0$, from the weakly interacting regime far from resonance
up to the unitary limit, $|a|\to\infty$. We will concentrate on the normal-fluid phase, 
i.e., on temperatures above the critical temperature $T_C$ of the 
superfluid-normal phase transition. The experimental data taken at 
these temperatures still show a rather strong damping of the collective 
modes, since the experiments are in fact neither in the collisionally 
hydrodynamic nor in the collisionless limit. This intermediate regime has
been studied theoretically in the framework of the Boltzmann equation 
(sometimes also called Boltzmann-Vlasov or Landau-Vlasov equation). In the
framework of the Boltzmann equation, interactions between the atoms
enter in two different places: On the one hand, each particle moves in
a mean field characterizing the average of the interaction with all
other particles. On the other hand, the particles undergo two-body
collisions, which are determined by the scattering cross-section.

In some of the early literature on this subject \cite{MenottiPedri,
PedriGueryOdelin, Toschi2003, Toschi2004,MassignanBruun}, the
mean field effects were considered within the Hartree
approximation. However, the most recent work by Bruun et al.
\cite{RiedlBruun, BruunSmithScissors} concentrates mainly on the
strongly interacting regime, where the Hartree approximation breaks
down: While more and more sophisticated models of the collision term
were developed, mean field effects were completely neglected
\cite{RiedlBruun, BruunSmithScissors}. Therefore, the predicted 
frequencies in the collisionless limit are those of an ideal gas, 
and in the hydrodynamic limit, the equation of state is that of an 
ideal gas, too. However, at least the experiment of \Ref{Altmeyer} 
clearly shows the importance of the mean field shift in the 
collisionless regime.

The aim of the present paper is to include mean field like effects on
the propagation of the particles into the Boltzmann equation in a way
which is appropriate also for the strongly interacting case. Our approach
is based on the in-medium scattering amplitude ($T$ matrix),
calculated in ladder approximation. This allows for a unified
description of mean field like effects on the propagation of the
particles (self-energy) and the modified cross-section entering the
collision term of the Boltzmann equation. We point out that in order
to have a consistent theory, it is important to use in the description
of the collective modes, which are small variations around
equilibrium, the same self-energy as in the calculation of the
equilibrium density profile. This can be seen, e.g., by looking at the
frequency of the sloshing mode, whose frequency, according to the Kohn
theorem \cite{Kohn, Brey}, must be equal to the corresponding trap
frequency.

Our paper is organized as follows. In \Sec{sec:interaction}, 
we concentrate on the description of the system in
equilibrium. After illustrating the breakdown of the Hartree
approximation, we review the treatment of the interacting
Fermi gas in the framework of the ladder approximation, introducing
the in-medium scattering amplitude, the single-particle self-energy,
and the in-medium cross-section. We discuss the quasiparticle
approximation which is necessary for describing the system in the
framework of the Boltzmann equation and which allows us to define a
quantity similar to the mean field in Hartree approximation. We show
the resulting density profiles and discuss the limits of validity of
the quasiparticle approximation. In \Sec{sec:modes}, we turn to the
description of collective modes in the framework of the Boltzmann
equation, which now includes the ``mean field'' and the in-medium
cross section. Using the standard method of taking moments of the
Boltzmann equation in phase space, we obtain semi-analytic expressions
for the collective mode frequencies which depend only on integrals of
equilibrium quantities. In \Sec{sec:resu}, we discuss our numerical
results for the scissors, radial quadrupole, and breathing modes and 
compare them with the available experimental data for temperatures above $T_C$. 
Finally, in \Sec{sec:conclusions}, we draw our conclusions and
give an outlook to further developments.

Throughout the paper, we will use in the derivations units with $\hbar
= k_B = 1$ ($\hbar = $ reduced Planck constant, $k_B$ = Boltzmann
constant).

\section{In-medium effective interaction}
\label{sec:interaction}
\subsection{Breakdown of the Hartree approximation}
\label{sec:Hartree}
Let us consider a two-component ($\uparrow$, $\downarrow$) uniform
Fermi gas in the normal phase. As long as the range of the interaction
is small compared with the mean distance between atoms, one can assume
a zero-range ($\delta$ function) interaction between atoms of opposite
spin, and the hamiltonian reads (in second quantization)
\begin{equation}
H = \int d^3r \Big(-\psi^\dagger \frac{\nabla^2}{2m} \psi + g
  \psi^\dagger_\downarrow \psi^\dagger_\uparrow \psi_\uparrow
  \psi_\downarrow\Big)\,,
\end{equation}
where $m$ and $g$ denote the atom mass and the coupling constant and
$\psi$ is the fermion field operator. We assume that the interaction
is attractive, i.e., $g < 0$. The coupling constant is related to the
atom-atom scattering length $a$ via
\begin{equation}
g = \frac{4\pi a}{m}\,.
\end{equation}
In Hartree approximation, the single-particle energies of $\uparrow$
and $\downarrow$ particles are shifted by $U_{H\uparrow} =
g\rho_\downarrow$ and $U_{H\downarrow} = g\rho_\uparrow$,
respectively. The exchange or Fock term vanishes since the interaction
is only between atoms with opposite spin. From now on, we will assume
that both spin states are equally populated, and we denote by $\rho =
\rho_\uparrow = \rho_\downarrow$ the density per spin state. Then the
Hartree shift is the same for both spin states, and we may write
\begin{equation}
U_H = g\rho\,.
\label{SigmaHartree}
\end{equation}

Let us now calculate the density as a function of the chemical
potential $\mu$. For our purposes it is enough to consider the
zero-temperature limit, i.e., temperatures $T\ll \epsilon_F$, where
\begin{equation}
\epsilon_F = \frac{k_F^2}{2m}\quad\mbox{and}\quad
  k_F = (6\pi^2\rho)^{1/3}
\label{epsilonFkF}
\end{equation}
are the Fermi energy and Fermi momentum, respectively. [Note that
\Eq{epsilonFkF} \textit{defines} $\epsilon_F$ and $k_F$ for a uniform
system, i.e., it remains valid even if the occupation numbers do not
resemble a step function because of temperature or correlation
effects. For a trapped system, however, we will use a different
definition of $\epsilon_F$ and $k_F$, see \Sec{sec:LDA}.] At zero
temperature, the relation between $\epsilon_F$ and $\mu$ is given by
$\epsilon_F = \mu-U_H$. Substituting \Eq{SigmaHartree} into this
relation, one obtains the following cubic equation for $k_F$:
\begin{equation}
-\frac{2ak_F^3}{3\pi m}-\frac{k_F^2}{2m}+\mu = 0\,.
\end{equation}
It is easy to see that this equation does not have a solution if $\mu$
exceeds a critical value given by $\mu_{\mathit{max}} =
\pi^2/(24ma^2)$, corresponding to a maximum density of
$\rho_{\mathit{max}} = \pi/(48|a|^3)$. The same value was found in
\Ref{Houbiers} as the density where the system becomes unstable
against separation into a low-density (gas) and a high-density (solid)
phase.

If the above arguments were correct, a low-temperature Fermi gas with
attractive interaction should be unstable as soon as $k_F |a| >
\pi/2$. However, we know from experiments that ultracold Fermi gases
are stable throughout the BCS-BEC crossover, including the unitary
limit $k_F |a|\to \infty$, because the system prefers to form pairs
instead of separating into two phases \cite{Nozieres}. The instability
is simply an artefact of the Hartree approximation and not physical.
\subsection{$T$ matrix}
\label{sec:Tmatrix}
In the Hartree approximation as described in the previous subsection,
the coupling constant $g$ was related to the scattering length $a$ in
free space. This means that it was implicitly assumed that the
scattering amplitude itself is the same in the gas as in free
space. As we will see, this assumption is the origin of the unphysical
instability of the Hartree approximation at high density or strong
interaction. For instance, as pointed out in \Ref{Heiselberg} the
scattering amplitude becomes proportional to $1/k_F$ instead of $a$ at
high density.

The approximation scheme we adopt here in order to calculate the
in-medium scattering amplitude is based on the non self-consistent
$T$ matrix approximation. In this approximation, the $T$ matrix is
given by the resummation of ladder diagrams, and it depends only on
the total energy $E = \omega+2\mu$ and the total momentum $\vek{k}$ of
the two atoms:
\begin{equation}
\Gamma(\omega,\vek{k}) = \frac{g}{1-g J(\omega,\vek{k})}\,,
\label{Tmatrix}
\end{equation}
where $J$ denotes the non-interacting two-particle Green's
function. Within the imaginary-time (Matsubara) formalism
\cite{FetterWalecka}, the latter is given by
\begin{multline}
J(i\omega_N,\vek{k})=-T \int
\frac{d^3 q}{(2 \pi)^3}\sum_{n~\mathit{odd}}
  \mathcal{G}_0(\omega_n,\vek{k}/2-\vek{q})\\
  \times\mathcal{G}_0(\omega_N-\omega_n,\vek{k}/2+\vek{q})\,,
\end{multline}
where $\omega_N$ and $\omega_n$ are, respectively, bosonic and
fermionic Matsubara frequencies, and $\mathcal{G}_0(\omega_n,\vek{k})
= 1/(i\omega_n-\xi^0_{\vek{k}})$ is the free (Matsubara) Green's
function, $\xi^0_{\vek{k}} = k^2/(2m)-\mu$ being the free
single-particle energy. After evaluation of the sum over $n$, the
retarded function $J(\omega,\vek{k})$ is obtained as usual by
analytic continuation. The result reads
\begin{equation}
J(\omega,\vek{k}) = \int \frac{d^3q}{(2\pi)^3}
  \frac{1-n^0_{\vek{k}/2+\vek{q}}-n^0_{\vek{k}/2-\vek{q}}}
  {\omega-\xi^0_{\vek{k}/2+\vek{q}}-\xi^0_{\vek{k}/2-\vek{q}}+i\eta}\,,
\label{Jdivergent}
\end{equation}
where $n^0_{\vek{k}} = 1/[\exp(\beta \xi^0_{\vek{k}})+1]$, with $\beta =
1/T$.

The problem with \Eqs{Tmatrix}--(\ref{Jdivergent}) is that $J$ is
divergent. To resolve this problem, one can introduce a momentum
cut-off $\Lambda$, determine the coupling constant $g$ as a function
of $\Lambda$ such that one recovers the correct scattering length $a$
in free space, and finally take the limit $\Lambda\to\infty$ keeping
the free-space scattering length $a$ constant \cite{Pieri}. In
this way one obtains for the $T$ matrix in free space
\begin{equation}
\Gamma_0(E,\vek{k}) = \frac{4\pi a}{m}
  \frac{1}{1+iaq_\mathit{cm}}\,,
\end{equation}
where $q_\mathit{cm} = \sqrt{mE-k^2/4}$ is the on-shell momentum in
the center-of-mass (CM) frame. If we now decompose $J$ into the
two-particle Green's function in free space, $J_0$, and a medium
correction, $\tilde{J}$, such that $J = J_0+\tilde{J}$, we may write
\begin{equation}
\Gamma(\omega,\vek{k}) = \frac{4\pi a}{m}
  \frac{1}{1+iaq_\mathit{cm}-\frac{4\pi a}{m}\tilde{J}}\,,
\end{equation}
with $q_\mathit{cm} = \sqrt{m(\omega+2\mu)-k^2/4}$. Even without
cut-off, the medium contribution $\tilde{J}$ is finite and given by
\begin{equation}
\tilde{J}(\omega,\vek{k}) = -\int \frac{d^3q}{(2\pi)^3}
  \frac{n^0_{\vek{k}/2+\vek{q}}+n^0_{\vek{k}/2-\vek{q}}}
  {\omega-\xi^0_{\vek{k}/2+\vek{q}}-\xi^0_{\vek{k}/2-\vek{q}}+i\eta}\,.
\label{jtilde}
\end{equation}
The imaginary part of $\tilde{J}$ can be given in closed form:
\begin{equation}
\Im \tilde{J}(\omega,\vek{k}) = \frac{m^2T}{2\pi k}
  \ln\Big(\frac{1+e^{-\beta\xi^0_-}}{1+e^{-\beta\xi^0_+}}\Big)\,,
\end{equation}
where $\xi^0_\pm = (k/2\pm q_\mathit{cm})^2/(2m)-\mu$. The real part is then
computed numerically via a dispersion relation,
\begin{equation}
\Re \tilde{J}(\omega,\vek{k}) = -\mathcal{P}\int
  \frac{d\omega^\prime}{\pi}
  \frac{\Im\tilde{J}(\omega^\prime,\vek{k})}
  {\omega-\omega^\prime}\,.
\label{dispersionrelation}
\end{equation}

As a by-product, the in-medium $T$ matrix allows us to determine the
critical temperature $T_C$ of the system, i.e., the temperature below
which the system becomes superfluid. As realized by Nozi\`eres and
Schmitt-Rink in their pioneering paper \cite{Nozieres}, the Thouless
criterion which relates $T_C$ to the temperature where the $T$ matrix
develops a pole at the Fermi level (i.e., at $\omega = 0$), remains
true at all couplings. Since the pole always appears first at
$\vek{k}=0$, the critical temperature can be obtained from the
equation
\begin{equation}
\Re\tilde{J}(\omega=0,\vek{k}=0;T=T_C)=\frac{m}{4\pi a}\,.
\label{Thouless}
\end{equation}
%

\subsection{Self-energy}
\label{sec:selfenergy}
Contrary to the zero-range interaction used in \Sec{sec:Hartree}, the
in-medium vertex function $\Gamma$ is now momentum and energy
dependent. This complicates the calculation of the single-particle
energy shift. Even the concept of such an energy shift may be
questioned if there are no well defined quasiparticles, as it is the
case in the ``pseudogap regime'' \cite{Perali}. In any case, the
appropriate object to calculate is the single-particle self-energy
$\Sigma$, which is well-defined and does not rely on the existence of
quasiparticles. The ladder self-energy can be written within the
Matsubara formalism as
\begin{multline}
\Sigma(i\omega_n,\vek{k}) = \int \frac{d^3p}{(2\pi)^3} 
T \sum_{n^\prime~\mathit{odd}} 
  \mathcal{G}_0(\omega_{n^\prime},\vek{k})\\
  \times \Gamma(i\omega_n+i\omega_{n^\prime}, \vek{p}+\vek{k})\,.
\end{multline}
Using analytic continuation to real energies, we find for the
imaginary part of the retarded self-energy:
\begin{multline}
\Im \Sigma(\omega,\vek{k}) = \int \frac{d^3 p}{(2\pi)^3}
  \Big(n^0_{\vek{p}}+\frac{1}{e^{\beta(\omega+\xi^0_{\vek{p}})}-1}
  \Big)\\ \times
  \Im\Gamma(\omega+\xi^0_{\vek{p}},\vek{k}+\vek{p})\,,
\end{multline}
which has to be evaluated numerically and from which the real part can
be obtained by a dispersion relation analogous to \Eq{dispersionrelation}.

Since we are going to use the Boltzmann equation for the description
of collective modes, we are implicitly assuming that the
quasi-particles, especially near the Fermi surface, are well defined,
which is of course a limitation of the range of applicability of our
approach \cite{BruunSmithShearViscosity}. If there are well-defined
quasiparticles, this means that their dispersion relation
$\xi_{\vek{k}}$, determined by the poles of the single-particle
Green's function, can be obtained from
\begin{equation}
\xi_{\vek{k}} = \xi^0_{\vek{k}} + \Re \Sigma(\xi_{\vek{k}},\vek{k})\,.
\end{equation}
Such a treatment is probably desirable but beyond the scope of the
present work. Here, we will completely neglect any energy and momentum
dependence of the self-energy. Since we are mainly interested in
momenta around the Fermi momentum, and hence energies around the Fermi
energy, we will make the approximation
\begin{equation}
\xi_{\vek{k}} \simeq \xi^0_{\vek{k}} + U\,,\quad \mbox{with}\quad
  U = \Re \Sigma(0,k_\mu)\,,
\label{spenergy}
\end{equation}
where $k_\mu = \sqrt{2m\mu}$ [which can actually be quite different
from $k_F$ as defined in \Eq{epsilonFkF}]. It is the quantity $U$
which will take the role of the mean field potential in the Boltzmann
equation.

\subsection{Density}
\label{sec:density}

The fact that the particles are interacting among each other changes
strongly the equation of state of the system, i.e., the relation
between the chemical potential $\mu$ and the density $\rho$. Within
the Hartree approximation, the relation $\rho(\mu)$ can trivially be
obtained, whereas in ladder approximation, the calculation of the
density for a given chemical potential $\mu$ is more involved.

In principle, the density can be obtained as
\begin{equation}
\rho = \int \frac{d^3p}{(2\pi)^3} T \sum_{n~\mathit{odd}}
  \mathcal{G}(\omega_n,\vek{p})
\label{rhoexact}
\end{equation}
where $\mathcal{G}$ denotes the full single-particle Green's function,
which according to Dyson's equation is given by $\mathcal{G}^{-1} =
\mathcal{G}_0^{-1} - \Sigma$\,. Here we will restrict ourselves to an
expansion of $\mathcal{G}$ up to first order in $\Sigma$, i.e., we
write
\begin{equation}
\mathcal{G} \simeq \mathcal{G}_0 + \mathcal{G}_0\Sigma\mathcal{G}_0\,.
\end{equation}
This leads us to the following expression for the density:
\begin{equation}
\rho = \rho_0 + \rho_1\,,
\end{equation}
where the free (uncorrelated) part, $\rho_0$, is given by
\begin{equation}
\rho_0 = \int \frac{d^3p}{(2\pi)^3} T \sum_{n~\mathit{odd}}
  \mathcal{G}_0(\omega_n,\vek{p})\\
  = \int \frac{d^3p}{(2\pi)^3} n^0_{\vek{p}}\,,
\end{equation}
while, after a lengthy calculation, the expression for the correlated
part, $\rho_1$, can be written as
\begin{multline}
\rho_1 = \int \frac{d^3k}{(2\pi)^3} T
    \sum_{n~\mathit{odd}} \mathcal{G}_0^2(\omega_n,\vek{k})
    \Sigma(i\omega_n,\vek{k})\\
  = \int \frac{d^3k}{(2\pi)^3} \int
    \frac{d\omega}{2\pi}
    \frac{1}{e^{\beta\omega}-1}\frac{d}{d\mu}\Im
      \ln[-\Gamma(\omega,\vek{k})]\,.
\label{rho1nsr}
\end{multline}
This is actually the result for the density initially given by
Nozi\`eres and Schmitt-Rink (NSR) in \Ref{Nozieres} and by S{\'a} de
Melo, Randeria, and Engelbrecht \cite{SadeMelo}.

\begin{figure}
\includegraphics[width=7cm]{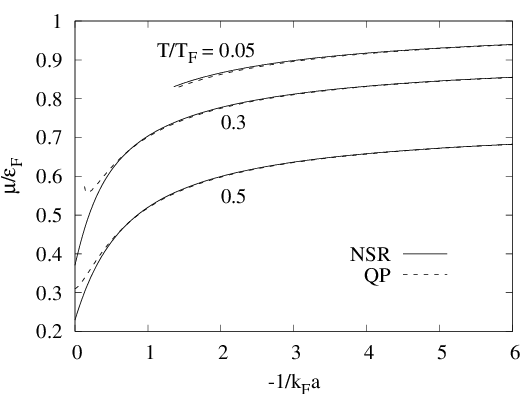}
\caption{\label{fig:muef} The ratio $\mu/\epsilon_F$, with $\mu$
  calcultated with the NSR density and the QP density, is plotted as a
  function of $-1/k_Fa$ for different temperatures: $T/T_F=0.05$,
  $T/T_F=0.3$ and $T/T_F=0.5$. The curve for $T/T_F = 0.05$ stops at
  $1/k_F a \approx 1.36$, since at this interaction strength the
  critical temperature is reached.}
\end{figure}

Unfortunately, the density formula given above is not suitable for
being used as the ground-state density in the Boltzmann equation. In
the Boltzmann equation, the density is expressed as an integral over
the distribution function $f$:
\begin{equation}
\rho = \int d^3p f(\vek{p})
\label{rhof}
\end{equation}
and $f$ must reduce to a simple Fermi function. Our quasiparticle(QP)
approximation for the density consists in using the single-particle
energy $\xi_{\vek{p}}$ as defined by \Eq{spenergy} in the calculation
of the occupation numbers, i.e.,
\begin{equation}
f_\eq(\vek{p}) = \frac{1/A}{e^{\beta \xi_{\vek{p}}}+1}\,,
\label{rho1qp}
\end{equation}
where we have introduced the abbreviation $A = (2\pi)^3$.  In this way
we obtain an alternative method for calculating the density as a
function of $\mu$. Fortunately, it turns out that in most cases both
ways of calculating the density give similar results. In order to
demonstrate this, we show in \Fig{fig:muef} 
the dimensionless ratio $\mu/\epsilon_F$ as a function of the
parameter $1/(k_F a)$ defining the interaction strength for various
temperatures (in units of the Fermi energy). One can see that the
interactions lead to a reduction of $\mu$ with respect to the ideal
gas result. At weak interactions, NSR and QP curves are in perfect
agreement, but in the strongly interacting case there can be
noticeable differences, especially at low temperatures close to $T_C$,
indicating the breakdown of the QP approximation. This is not
surprising, since in this case the system is in the ``pseudogap
regime'' \cite{Perali}.

\subsection{Cross section}
\label{sec:xsection}
The interaction between atoms is not only responsible for the
single-particle energy shift. It also determines the collision rate of
the atoms, which will play a central role for the properties of
collective modes. The important quantity is the cross section
$\sigma$. In the case of a zero-range $s$-wave interaction, the cross
section for two atoms in free space with momenta $\vek{p}_1$ and
$\vek{p}_2$ before the collision and $\vek{p}^\prime_1$ and
$\vek{p}^\prime_2$ after the collision is given by
\cite{Landau3}
\begin{equation}
\frac{d\sigma_0}{d\Omega} = \frac{a^2}{1+(qa)^2}
\label{freexsection}
\end{equation}
where $\vek{q} = (\vek{p}_1 - \vek{p}_2) / 2$ is the incoming momentum
in the CM frame and $\Omega$ is the solid angle after the scattering,
i.e., $d\Omega = 2\pi\sin\theta d\theta$, where $\theta$ is the angle
between $\vek{q}$ and $\vek{q}^\prime = (\vek{p}^\prime_1 -
\vek{p}^\prime_2) / 2$ (note that $|\vek{q}| = |\vek{q}^\prime|$
because of energy and momentum conservation).

As pointed out in \Refs{RiedlBruun,BruunSmithShearViscosity}, the
cross section is strongly modified by medium effects. In terms of the
$T$ matrix, the in-medium cross section can be written as
\begin{equation}
\frac{d\sigma}{d\Omega} = \Big|\frac{m}{4\pi}
  \Gamma\Big(\frac{k^2}{4m}+\frac{q^2}{m}-2\mu,\vek{k}\Big)
  \Big|^2
\label{xsection}
\end{equation}
which now depends also on the total momentum $\vek{k} = \vek{p}_1 +
\vek{p}_2 = \vek{p}^\prime_1 + \vek{p}^\prime_2$ of the two atoms.

\begin{figure}
\includegraphics[width=7cm]{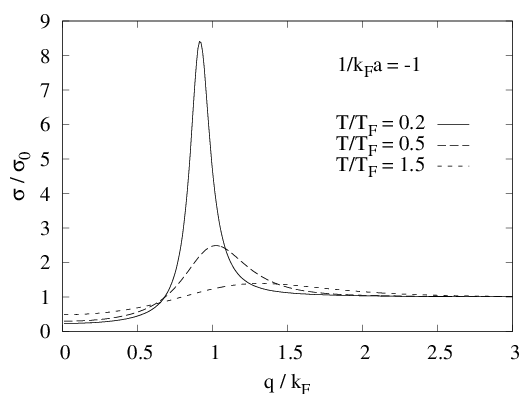}
\caption{\label{fig:xsection} The ratio of the in-medium cross section
  and the vacuum one is displayed for total momentum $k=0$ as a
  function of the relative momentum $q$. The results are shown for
  different temperatures for fixed interaction strength $k_F a =
  -1$.}
\end{figure}

It is not easy to see a priori what will be the effect of these medium
modifications, since depending on $k$ and $q$ the cross section can be
enhanced or reduced as compared with the cross section in free
space. This is shown in \Fig{fig:xsection}
where we display the cross section as a function of $q$ for the case
$k = 0$ (where the medium effect is supposed to be strongest) for
various temperatures for fixed interaction strength $k_F a = -1$. The
strong enhancement of the cross-section near the critical temperature
is a precursor of the singularity of the $T$ matrix at the critical
temperature \cite{RiedlBruun,BruunSmithShearViscosity}.

Let us mention that the same effect has already been found some years
ago in the in-medium nucleon-nucleon cross section in low-temperature
nuclear matter \cite{AlmRoepke}.

\subsection{Local-density approximation}
\label{sec:LDA}
Until now we considered a uniform system where the atoms are not
trapped by an external potential. In order to include the trap
potential $V_T(\vek{r})$, we will make use of the local-density
approximation (LDA), where the system is treated as locally
homogeneous, with an $\vek{r}$ dependent chemical potential which is
given by
\begin{equation}
\mu(\vek{r}) = \mu_0-V_T(\vek{r}).
\label{LDA}
\end{equation}
This approximation should be valid as long as the potential varies
only slowly, i.e., on length scales which are large compared with
$1/k_F$. By the way, this condition is also necessary for the validity
of the Boltzmann equation which will be used later to describe the
collective modes.

Within the local-density approximation, all equilibrium quantities of
the system discussed in the preceding sections, like $\Gamma$, $U$,
$\rho$, $d\sigma/d\Omega$, etc., acquire an additional $\vek{r}$
dependence via the dependence of $\mu$ on $\vek{r}$.

In practical calculations, we will use a harmonic trap potential
\begin{equation}
V_T(\vek{r}) = \frac{m}{2}\sum_{i=x,y,z}\omega_i^2 r_i^2\,.
\end{equation}
Besides the fact that experimental traps are almost harmonic, the use
of a harmonic potential has the advantage that it is sufficient to
calculate the equilibrium quantities once for a spherical trap with
the same number of atoms and the average frequency $\bar{\omega} =
(\omega_x\omega_y\omega_z)^{1/3}$. Then, the equilibrium quantities in
the deformed trap can easily be obtained from the corresponding ones
in the spherical trap by the change of variables $\tilde{r}_i = r_i
\omega_i / \bar{\omega}$ (for $i = x,y,z$).

\begin{figure}
\includegraphics[width=7cm]{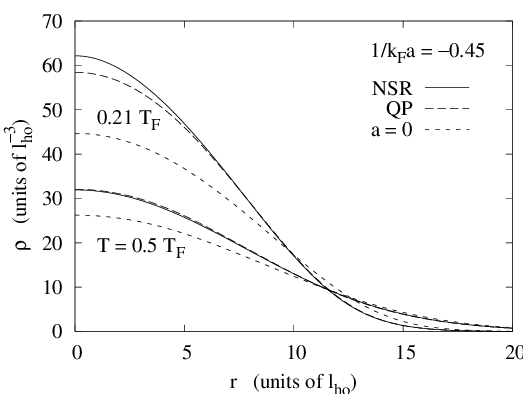}
\caption{\label{fig:density} Density profiles for 400000 atoms with
  $1/k_F a = -0.45$ at various temperatures, calculated using the NSR
  formula (solid lines) or the QP approximation (long dashes). For
  comparison, the corresponding density profiles of an ideal Fermi gas
  (short dashes) are shown, too. The length unit is the
  harmonic-oscillator length $l_{ho} = \sqrt{1/m\bar{\omega}}$.}
\end{figure}

As an example, we show in \Fig{fig:density}
the density profile of $N = 400000$ atoms in a trap with fixed
interaction strength $1/k_F a =-0.45$ at different temperatures. Note
that we follow here the usual convention of the experimental papers,
using for the trapped gas a definition of $k_F$ and $\epsilon_F$ which
is different from that used in the uniform case
[cf. \Eq{epsilonFkF}]. In the trapped case, $k_F$ and $\epsilon_F$
refer to the values of $k_F$ and $\epsilon_F$ in the center of the
trap calculated for an \textit{ideal} Fermi gas at \textit{zero
temperature}. Therefore, $\epsilon_F$ and $k_F$ are determined by the
number of atoms, $N$, and the average trap frequency as follows:
\begin{equation}
\epsilon_F = (3N)^{1/3} \bar{\omega}\,,
\quad k_F = \sqrt{2m\epsilon_F}\,.
\label{epsilonFkFtrap}
\end{equation}
One can see that the interaction leads to a noticeable change of the
density profile, especially at low temperature. The agreement between
the NSR and QP results is almost perfect in the case $T/T_F = 0.5$,
while at $T/T_F = 0.21$, corresponding to the critical temperature,
the QP density is too small, indicating again the breakdown of the QP
approximation in the pseudogap regime. However, it should be mentioned
that for $1/k_F a = -0.45$, one has to go very close to $T_C$ in order
to see this effect; it is more important in the unitary limit. Let us
mention that similar density profiles can be found in the literature,
even for temperatures below the critical one \cite{PeraliPieri}.

\section{\label{sec:modes}Collective Modes}
\subsection{\label{sec:LinBol}Linearized Boltzmann equation}
After the discussion of static properties, let us now turn to the
description of collective modes of a trapped Fermi gas. We remind the
reader of three assumptions mentioned earlier: (a) the density is
supposed to vary (due to the trap as well as the collective motion)
only on large length scales; (b) the temperature has to be above the
superfluid transition temperature $T_C$; (c) the quasiparticles near
the Fermi surface have to be well defined. In addition, the Boltzmann
equation is only valid if (d) the time dependence of the excitations
under consideration (in the case of the collective modes under
consideration this time scale is set by the the trap frequency) are
slow compared to the ``correlation time'' \cite{Danielewicz}. Under
these assumptions, the dynamics of the system can be described by the
semi-classical distribution function $f(\vek{r},\vek{p},t)$ whose time
evolution is governed by the Boltzmann equation \cite{Landau10}:
\begin{equation}
\dot{f}+ \dot{\vek{r}}\cdot\vek{\nabla}_r f+
\dot{\vek{p}}\cdot\vek{\nabla}_p f
=-I[f]~,
\label{eq:Boltz}
\end{equation}
where $\dot{\vek{r}}$ and $\dot{\vek{p}}$ satisfy the classical
equations of motion. In the case of a quasiparticle dispersion
relation as given by \Eq{spenergy} with an $\vek{r}$ dependent
chemical potential as given by \Eq{LDA}, the velocity and
acceleration read
\begin{gather}
\dot{\vek{r}}=\vek{\nabla}_p\xi_{\vek{p}} =
\frac{\vek{p}}{m}\,,\\
\dot{\vek{p}}=-\vek{\nabla}_r\xi_{\vek{p}} = -\vek{\nabla}_r(V_T+U)\,,
\label{eq:pdot}
\end{gather}
since, within LDA, $\xi_{\vek{p}} \to \xi_{\vek{p}}(\vek{r}) =
p^2/(2m)+V_T(\vek{r})+U(\vek{r})-\mu_0$. Note that there are two
sources of $\vek{r}$ dependence of the self-energy $U$. In
equilibrium, $U$ depends on $\vek{r}$ only via $\mu(\vek{r}) =
\mu_0-V_T(\vek{r})$. More generally, in particular out of equilibrium,
the self-energy depends on the distribution function $f$, i.e., we may
write $U = U[f]$.

The distribution function $f$ is related to the density per spin state
by \Eq{rhof} and we assume that, as in equilibrium, the distribution
functions for the two spin states are the same, i.e., $f_\downarrow =
f_\uparrow = f$. This is true if the trap potential and the excitation
operator of the collective mode are spin independent.

The functional $I[f]$ appearing on the rhs of \Eq{eq:Boltz} is the
collision integral. It describes collisions between atoms with
opposite spin and depends on the differential scattering cross section
as \cite{Landau10}
\begin{multline}
I[f]=\int d^3p_1\int d\Omega \frac{d\sigma}{d\Omega}
  |\vek{v}-\vek{v_1}|
  [f f_1 (1-Af^\prime) (1-Af_1^\prime)\\
  -f^\prime f_1^\prime (1-Af) (1-Af_1)]\,,
\end{multline}
where $\vek{p}$ and $\vek{p}_1$ are the incoming momenta,
$\vek{p}^\prime$ and $\vek{p}_1^\prime$ are the outgoing ones;
$\vek{v}$ and $\vek{v}_1$ are the incoming velocities $\vek{p}/m$ and
$\vek{p}_1/m$, respectively, $\Omega$ is the solid angle formed by the
incoming relative momentum $\vek{p}-\vek{p}_1$ and the outgoing
relative momentum $\vek{p}^\prime-\vek{p}_1^\prime$ of the two atoms,
$f = f(\vek{r},\vek{p},t)$, $f_1 = f(\vek{r},\vek{p}_1,t)$, $f^\prime
= f(\vek{r},\vek{p}^\prime,t)$, etc. The factors of the type $(1-A f)$
are absent in the classical Boltzmann equation. They are a consequence
of Fermi statistics and ensure that an atom cannot be scattered into a
state which is already occupied. This Pauli blocking effect can result
in a strong reduction of the collision rate, especially at low
temperatures.

In order to study the collective modes of the trapped gas, we consider
a small deviation $\delta f=f-f_\eq$ of the distribution
function from the equilibrium one. Usually $\delta f$ is strongly
peaked at the Fermi surface, but it can conveniently be written as
\begin{equation}
\delta f(\vek{r},\vek{p},t)=f_\eq(\vek{r},\vek{p})[1-A
f_\eq(\vek{r},\vek{p})] \Phi(\vek{r},\vek{p},t)\,,
\label{eq:deltaf}
\end{equation}
with a smooth function $\Phi$ \cite{Landau10}. Expanding the Boltzmann
equation (\ref{eq:Boltz}) to linear order in the deviations from
equilibrium, and considering that a change $\delta f$ of the
distribution function results in a change $\delta U$ of the
self-energy, we obtain
\begin{multline}
f_\eq(1-Af_\eq) \Big(\dot{\Phi} +
  \frac{\vek{p}}{m}\cdot\vek{\nabla}_r\Phi -
  \vek{\nabla}_r(V_T+U_\eq)\cdot\vek{\nabla}_p\Phi \\
  + \beta\frac{\vek{p}}{m}\cdot\vek{\nabla}_r\delta U
\Big) = -I[\Phi]
\label{eq:BoltzLin}
\end{multline}
with the linearized collision integral
\begin{multline}
I[\Phi] = \int d^3 p_1\int d\Omega
\frac{d\sigma}{d\Omega} |\vek{v}-\vek{v_1}|
  f_\eq f_{\eq\,1}\\
  \times (1-Af_\eq^\prime)(1-Af_{\eq\,1}^\prime)
  (\Phi+\Phi_1-\Phi^\prime-\Phi_1^\prime)\,.
\label{collisionterm}
\end{multline}
Now we have to specify $\delta U$, which appears in the lhs of
\Eq{eq:BoltzLin}. Since we neglect any possible momentum dependence of
$U$ and $\delta U$, it is clear that $\delta U$ can be written as
\begin{equation}
\delta U(\vek{r},t) = \int d^3p \gamma(p) \delta f(\vek{r},\vek{p},t) \,,
\end{equation}
where $\gamma(p)$ is the functional derivative $\delta U[f]/\delta f$,
taken at $f_\eq$. In the low-temperature limit, $\delta f$ is so
strongly peaked at the Fermi momentum $p_F$ that $\gamma(p)$ may be
replaced by a constant $\gamma_0 = \gamma(p_F)$, as in Fermi liquid
theory \cite{Landau9}. This results in $\delta U = \gamma_0\delta\rho$, with
\begin{equation}
\delta\rho(\vek{r},t) = \int d^3p \delta f(\vek{r},\vek{p},t)
  = \int d^3p f_\eq(1-A f_\eq)\Phi\,.
\end{equation}
In addition, by choosing a particular form for $\delta f$, namely
$\delta f = \partial f_\eq/\partial \mu$, we can identify $\gamma_0$
with the derivative $\partial U_\eq/\partial \rho_\eq$ taken at
constant $T$, i.e.,
\begin{equation}
\delta U(\vek{r},t) = \frac{\partial U_\eq}{\partial
  \rho_\eq}\Big|_{\rho_\eq(\vek{r}),T}
  \delta\rho(\vek{r},t)\,.
\label{deltaU}
\end{equation}
In the present work, we shall assume that \Eq{deltaU} is a reasonable
approximation also at higher temperatures, although it cannot be
rigorously justified in this case.

Equation (\ref{eq:BoltzLin}) together with (\ref{deltaU}) constitutes
the starting point for our study of collective modes with in-medium
effects. It is a generalization of the Boltzmann equation used in
\Refs{PedriGueryOdelin, Toschi2003, Toschi2004, MassignanBruun} in the case $U = g\rho$
(Hartree approximation).

\subsection{Trial function}
\label{sec:Phi}
As mentioned in \Sec{sec:LinBol}, the function
$\Phi(\vek{r},\vek{p},t)$ characterizing the deviation from
equilibrium can be supposed to be smooth in phase space. This allows
us to make a simple ansatz for $\Phi$ with a small number of
coefficients rather than solve the linearized Boltzmann equation
(\ref{eq:BoltzLin}) exactly.

For any collective mode of interest, the trial function
$\Phi(\vek{r},\vek{p},t)$ has to contain at least those terms which
are necessary to generate the velocity field $\vek{u}(\vek{r},t)$
characterizing the mode \cite{Khawaja}. The presence of a
velocity field modifies $f_{eq}$ into
\begin{equation}
f(\vek{r},\vek{p},t)=f_\eq[\vek{r},\vek{p}-m\vek{u}(\vek{r},t)]
\end{equation}
and leads to a deviation
\begin{equation}
\delta f \simeq -\beta f_\eq(1-Af_\eq)\,\vek{p}\cdot\vek{u}\,,
\end{equation}
i.e., the trial function $\Phi$ must at least contain a term
proportional to $\vek{p}\cdot\vek{u}$. When this term is inserted into
the linearized Boltzmann equation (\ref{eq:BoltzLin}), the operator
$\vek{p}/m \cdot \vek{\nabla}_r - \vek{\nabla}_r(V_T+U_\eq) \cdot
\vek{\nabla}_p$ on the lhs of \Eq{eq:BoltzLin} generates new terms, as
do the $\delta U$ term on the lhs and the collision term $I$ on the
rhs. In general, the number of terms is infinite and the system cannot
be closed.

However, in the case of an ideal gas ($U = \delta U = I = 0$) in a
harmonic potential $V_T$, and if $u$ is at most linear in the
coordinates, it is possible to solve \Eq{eq:BoltzLin} with a finite
number of terms. For instance, a term proportional to $xp_x$ generates
terms proportional to $x^2$ and $p_x^2$, and no other terms are
needed. In the opposite limit of an extremely strong collision term,
i.e., in the hydrodynamic regime, a linear velocity field solves
exactly the hydrodynamic equations if the equation of state can be
approximated by a polytropic one, which is in many cases an excellent
approximation \cite{Cozzini}. We therefore assume that also in our
case it will be a good approximation to include into $\Phi$ only those
terms which appear in the ideal gas case (of course the coefficients
will change).

To be explicit, we will focus on the scissors mode ($S$), the radial
quadrupole mode ($Q$), and the breathing modes ($B$). In order to
check the consistency of our model, we will also consider the Kohn
mode (center-of-mass or sloshing mode, $K$). The velocity fields and
the corresponding trial functions for these modes are given in
\Tab{tab:Phi}. Note that in the case of the breathing modes, the axial 
and the radial modes cannot be treated separately because they are coupled
(although the coupling may be weak in very elongated traps).
\begin{table*}
\caption{\label{tab:Phi} Velocity fields and corresponding ansatz
functions $\Phi$ for the different modes under consideration.}
\begin{ruledtabular}
\begin{tabular}{ccccc}
mode && trap frequencies & \vek{u}(\vek{r},t) & 
  $\Phi(\vek{r},t) e^{i\omega t}$\\
\hline
sloshing &(K)& arbitrary & $\propto (1,0,0) $ &
  $c_1 x + c_2 p_x$\\
scissors &(S)& $\omega_x > \omega_y \gg \omega_z$ &
  $\propto (y,-x,0)$ &
  $c_1 xy+c_2 x p_y + c_3 y p_x + c_4 p_x p_y$\\
radial quadrupole &(Q)& $\omega_x = \omega_y = \omega_r \gg \omega_z$ &
  $\propto (x,-y,0)$ &
  $c_1(x^2-y^2)+c_2(x p_x-y p_y)+c_3(p_x^2-p_y^2)$\\
$\left.\begin{array}{l}\mbox{radial}\\ \mbox{axial}\end{array}\right\}
  \mbox{breathing}$ &(B)& $\omega_x = \omega_y = \omega_r \gg \omega_z$ &
  $\begin{array}{c}\propto (x,y,0)\\ \propto(0,0,z)\end{array}$ &
  $\left\{\begin{array}{l}c_1(x^2+y^2)+c_2 z^2+c_3(x p_x+y p_y)\\
    +c_4 z p_z+c_5(p_x^2+p_y^2)+c_6 p_z^2\end{array}\right.$
\end{tabular}
\end{ruledtabular}
\end{table*}

\subsection{Frequency and damping of collective modes}
\label{sec:frdamp}
By inserting each trial function $\Phi$ into the linearized Boltzmann
equation (\ref{eq:BoltzLin}) and taking moments of the equation,
namely multiplying it by any of the terms contained in $\Phi$ and then
integrating over $\vek{r}$ and $\vek{p}$, one obtains a set of
homogeneous linear equations for the coefficients $c_i$. The condition
that the coefficient matrix determinant is zero yields an equation for
the frequencies of the collective mode.

Let us start by the center-of-mass oscillation of the cloud, known as
sloshing or Kohn mode ($K$). In experiments, this mode is used in
order to determine the trap frequency with high precision
\cite{Altmeyer2}, since it is known to be an undamped oscillation with
the frequency of the trap, independently of the interaction
\cite{Kohn,Brey}. It is an important test of the consistency of our
method to check that this property is preserved.

Multiplying \Eq{eq:BoltzLin} (with $\Phi = c_1 x + c_2 p_x$) by $x$
and $p_x$ and integrating over $\vek{r}$ and $\vek{p}$, we obtain the
following system of equations [note that the collision term
(\ref{collisionterm}) on the rhs of the Boltzmann equation
(\ref{eq:BoltzLin}) does not contribute since $I[x] = I[p_x] = 0$)]:
\begin{gather}
\frac{-i\omega}{m\omega_x^2}\Big(N_\uparrow-\frac{C}{3}\Big)c_1
  -N_\uparrow c_2 = 0\,,\\
\Big(N_\uparrow - \frac{C}{3}\Big) c_1 -i\omega mN_\uparrow c_2 = 0\,,
\end{gather}
where
\begin{equation}
N_\uparrow = \int d^3 \tilde{r}\, \rho_\eq
\end{equation}
denotes the number of atoms per spin state and
\begin{equation} C = \int d^3 \tilde{r}\,d^3p\,\beta f_\eq(1-Af_\eq)\,
  \tilde{r} \frac{\p U_\eq}{\p \tilde{r}}
\end{equation}
is a constant depending on the interaction. When calculating the
determinant, we obtain $\omega = \omega_x$, independently of the
interaction, as it should be. Of course, analogous results are
obtained for the sloshing modes in the $y$ and $z$ direction.

If we repeat the same steps as before for the case of the scissors,
quadrupole, or breathing mode, an additional complication arises from
the fact that now the collision term on the rhs of the linearized
Boltzmann equation (\ref{eq:BoltzLin}) gives a non-vanishing
contribution. More precisely, only the terms in $\Phi$ which are
quadratic in momentum contribute, since $I[r_ir_j] = I[r_ip_j] = 0$
for $i,j = x,y,z$. Using the symmetry properties of the explicit
expression for $I[p_i p_j]$, one can furthermore show that the $r_k
r_l$ and $r_k p_l$ moments of $I[p_ip_j]$ ($i,j,k,l = x,y,z$) vanish
and only moments involving two momenta,
\begin{equation}
I_{ijkl} = \int d^3r\, d^3p\, I[p_ip_j] p_k p_l\,,
\end{equation}
survive. Using $I[p^2] = 0$, one can show that these must be of the
form
\begin{equation}
I_{ijkl} = I_S \Big(\delta_{ik}\delta_{jl}+\delta_{il}\delta_{jk}
  -\frac{2}{3}\delta_{ij}\delta_{kl}\Big)\,,
\label{IS}
\end{equation}
where $I_S$ is the moment which is relevant for the scissors mode, i.e.,
\begin{equation}
I_S = I_{xyxy} = \int d^3r\, d^3p\,I[p_x p_y]p_xp_y\,.
\end{equation}
Some more details on how $I_S$ is calculated are given in Appendix
\ref{app:relaxation}. Now let us define the corresponding relaxation
time $\tau$ as
\begin{equation}
\frac{1}{\tau}=\frac{\int d^3\, d^3p\, I[p_xp_y]p_xp_y}
  {\int d^3r\,d^3p\,f_\eq(1-Af_\eq) p_x^2p_y^2} 
  = \frac{3\beta I_S}{m^2 E_\kin}\,,
\label{eq:tausci}
\end{equation}
where $E_\kin$ denotes the kinetic energy (cf. Appendix
\ref{app:virial}).
It is essentially this parameter which governs the temperature
dependence of the mode frequencies and damping rates. The definition
(\ref{eq:tausci}) is identical with that introduced in
\Ref{RiedlBruun}.

Using the definition (\ref{eq:tausci}), we can write the equation for
the scissors mode frequencies in the following form (see Appendix
\ref{app:scissors} for more details):
\begin{equation}
\frac{i\omega}{\tau}(\omega^2-\omega_{S,h}^2)+(\omega^2-\omega_{S,c+}^2)
  (\omega^2-\omega_{S,c-}^2)=0,
\label{SciFrDa}
\end{equation}
where $\omega_{S,h}$ and $\omega_{S,c\pm}$ are the frequencies in the
hydrodynamic ($\omega\tau\rightarrow 0$) and collisionless
($\omega\tau\rightarrow\infty$) limits, respectively. The hydrodynamic
frequency is given by
\begin{equation}
\omega_{S,h}^2=\omega_x^2+\omega_y^2\,,
\label{eq:whSciMF}
\end{equation}
and does not depend on the interaction. In the collisionless limit,
there are two modes with different frequencies, corresponding to
rotational [$\vek{u} \propto (y,-x,z)$] and irrotational [$\vek{u}
\propto (y,x,0)$] velocity fields. In a non-interacting gas, these two
modes have the frequencies $\omega_{S,\cl\pm} = \omega_x \pm
\omega_y$. In the interacting case, they are changed to
\begin{multline}
\omega_{S,\cl\pm}^2 = (\omega_x^2+\omega_y^2)(1-\chi/2)\\
  \pm\sqrt{4\omega_x^2\omega_y^2(1-\chi+\chi^2/8)+(\omega_x^4+\omega_y^4)
  \chi^2/4}\,,
\label{eq:wcSciMF}
\end{multline}
where $\chi$ is the interaction dependent parameter defined in
\Eq{eq:chi}.

For the radial quadrupole mode, the equation for the frequencies has
the form
\begin{equation}\label{QuaFrDa}
i\omega(\omega^2-\omega_{Q,\cl}^2)-\frac{1}{\tau}(\omega^2-\omega_{Q,\hd}^2)
  =0\,,
\end{equation}
the hydrodynamic frequency is, again, independent of the interaction
and given by
\begin{equation}
\omega_{Q,\hd}^2 = 2\omega_r^2\,,
\end{equation}
while the frequency in the collisionless limit depends on the
interaction:
\begin{equation}
\omega_{Q,\cl}^2 = 4 \omega_r^2 (1-\chi/2)\,.
\end{equation}
In the case $U = g\rho$ (Hartree approximation), $\chi$ reduces to
$3E_\inter/2E_\pot$, where $E_\inter$ and $E_\pot$ denote the
interaction and potential energies (cf. Appendix \ref{app:virial}). In
this case, our limiting frequencies (in the hydrodynamic and
collisionless limits: $\omega\tau\to 0$ and $\omega\tau\to \infty$)
agree with those of \Ref{MenottiPedri, MassignanBruun}.

In the case of the breathing mode, we obtain two frequencies
$\omega_{B\pm}$, corresponding to the axial and radial breathing
modes.  The low-lying mode ($\omega_{B-}$) corresponds essentially to
a motion in the $z$ direction (axial breathing mode), while the
high-lying mode ($\omega_{B+}$) corresponds to a motion in the radial
direction (radial breathing mode). The equation for the frequencies
has the form
\begin{multline}
i\omega(\omega^2-\omega_{B,\cl+}^2) (\omega^2-\omega_{B,\cl-}^2) \\ -
\frac{1}{\tau} (\omega^2 - \omega_{B,\hd+}^2)
(\omega^2-\omega_{B,\hd-}^2)=0\,.
\label{BreFrDa}
\end{multline}
The expressions for the limiting frequencies $\omega_{B,\hd\pm}$ and
$\omega_{B,\cl\pm}$ are given in Appendix \ref{app:breathing}.
%

\section{\label{sec:resu}Results and discussion}
%
In this section we will present our numerical results for the
scissors, radial quadrupole, and radial breathing modes. We will
discuss the frequencies and damping rates as functions of the
temperature for finite values of $1/k_F a$ and for the unitary limit
($1/k_F a = 0$). The frequencies $\omega$ and damping rates $\Gamma$
are determined by the real and imaginary parts of the solutions of
\Eqs{SciFrDa}, (\ref{QuaFrDa}), and (\ref{BreFrDa}), respectively.

So far, most experiments have been done on resonance, i.e., for $1/k_F
a = 0$. Two exceptions are the study of the radial quadrupole mode
over the whole crossover region by Altmeyer et al. \cite{Altmeyer} and
of the scissors mode at $1/k_Fa = -0.45$ by Wright et
al. \cite{Wright}. In addition to these two experiments, we will
compare our results to the scissors, quadrupole, and breathing mode
experiments at unitarity described in \Refs{Wright,RiedlBruun}.
%
\subsection{Radial quadrupole mode at $1/k_Fa = -1.34$}
\label{sec:quafin}
%
In the first experiment on the radial quadrupole mode on the BCS side
of the BEC-BCS crossover \cite{Altmeyer}, the trap had frequencies
$\omega_r = 2\pi\times 370$ Hz and $\omega_z = 2\pi\times 22$ Hz and
contained $N = 400000$ $^6$Li atoms. The highest magnetic field used
in this experiment, corresponding to the weakest interaction, resulted
in $1/k_F a = -1.34$, which is the value we will consider here. The
temperature is unfortunately not known, but we assume that it was
between $0.03 T_F$, the lowest value ever reported by the Innsbruck
group \cite{Bartenstein}, and $0.1 T_F$, the upper value given in
\Ref{Altmeyer}.

\begin{figure*}
\includegraphics[width=5.8cm]{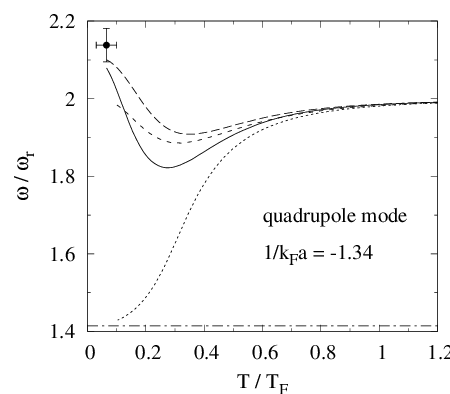}
\includegraphics[width=5.8cm]{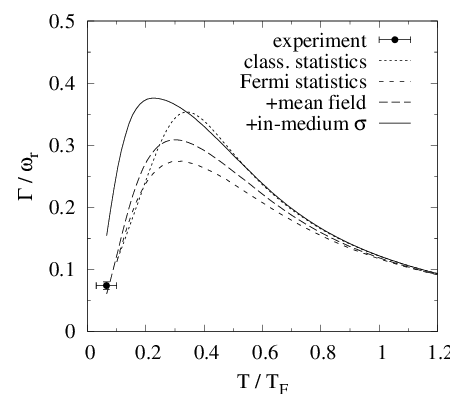}
\includegraphics[width=5.8cm]{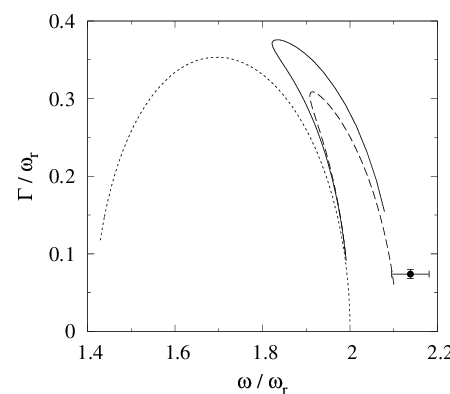}
\caption{\label{fig:quafin} Frequency and damping of the quadrupole
 mode for $1/k_F a = -1.34$. The experimental result is taken from
 \Ref{Altmeyer}. The three panels display (from left to right)
 frequency as function of temperature, damping as function of
 temperature, and damping vs. frequency. The different lines represent
 different levels of sophistication of the calculation: Starting from
 a calculation for a classical gas without any mean field and with the
 scattering cross-section in vacuum (dotted lines), we include the
 Pauli principle in the equilibrium density profile and in the
 collision integral (short dashes), then on top of that the mean field
 $U$ (long dashes), and finally also the in-medium scattering cross
 section (solid line). The dash-dotted line represents the
 hydrodynamic frequency, $\omega_{Q,\hd} = \sqrt{2}\omega_r$.}
\end{figure*}

Results for frequency and damping as functions of temperature are
shown in the first two panels of \Fig{fig:quafin},
while the third panel shows the damping rate vs. the frequency. This
latter representation was proposed in \Ref{RiedlBruun} in order to get
rid of the temperature, which cannot easily be experimentally
determined.  The single data point shows unambiguously the necessity
of the inclusion of the mean field since the measured frequency ($\sim
2.1\omega_r$) lies clearly above the limiting value for a
collisionless gas without mean field ($2\omega_r$).

The theoretical curves shown in \Fig{fig:quafin} represent different
levels of approximation in the calculation. In order to see the effect
of the different improvements of the theory, we include them one after
another. We start with a classical Fermi gas (dotted lines), using
Boltzmann distribution functions $(f_\eq =
e^{-\beta(p^2/2m+V_T-\mu_0)})$ in the calculation of the density
profile, without any mean field effects and with the free
cross-section, \Eq{freexsection}, without Pauli-blocking factors
$(1-Af_\eq)$ in the collision term. Within this approximation, the
system shows hydrodynamic behavior ($\omega \to \sqrt{2}\omega_r$) at
low temperature and collisionless behavior ($\omega\to 2\omega_r$) at
high temperature, with strong damping $\Gamma$ in the intermediate
regime. In the representation of $\Gamma$ vs. $\omega$, this results
in a curve similar to a semi-circle. The hydrodynamic behavior at low
temperature is of course an artefact of neglecting the Pauli blocking
in the collision term and it is in clear contradiction to the measured
frequency.

In order to cure this problem, we include the effect of Fermi
statistics (short dashes), i.e., we use the Fermi distribution
function $f_\eq = 1/(e^{\beta(p^2/2m+V_T-\mu_0)}+1)$ in the
calculation of the density profile, and the Pauli-blocking factors
$(1-Af_\eq)$ in the collision term. At this stage, we still use the
free cross section and we do not include any mean field. Due to the
Pauli blocking factors, the collision rate goes now to zero at low
temperature, and therefore the system approaches the collisionless
frequency $2\omega_r$ in both the low and high temperature
limits. The highest damping, and as a consequence the lowest
frequency, is reached at a temperature of $\sim 0.3 T_F$. Since no
mean field is included, $\omega$ and $\Gamma$ depend only on a single
parameter, namely on $\tau$ [cf. \Eq{QuaFrDa}]. Therefore the results
lie on the same curve in the $\omega-\Gamma$ plane as in the case of a
classical gas (dotted curve), but this time only the small part of the
curve corresponding to large values of $\tau$ is covered. Although the
frequency at low-temperature is now in better agreement with the data,
it is still too low, since we have not yet included the mean field.

The third step consists in switching on the mean field $U$ (long
dashes), i.e., the density profiles are now calculated with $f_\eq =
1/(e^{\beta(p^2/2m+V_T+U-\mu_0)}+1)$ and the $U_\eq$ and $\delta U$
terms are included in the Boltzmann equation (\ref{eq:BoltzLin}). We
still keep the free scattering cross section. Our calculation is
limited to temperatures above $0.06 T_F$, corresponding to the
critical temperature of the system. The mean field does not have a
dramatic effect on the damping, but it increases the frequency,
especially at low temperature. Frequency and damping at the lowest
temperature are now both in excellent agreement with the measured
values. Note that the inclusion of the mean field modifies
qualitatively the curve $\Gamma$ vs. $\omega$ shown in the third panel
of \Fig{fig:quafin}.

Finally, we replace the free scattering cross section by the in-medium
one, \Eq{xsection} (solid lines). Unfortunately, the good agreement
between the theoretical results and the measured frequency and damping
at low temperature is deteriorated: The resulting damping is too high
by a factor of two and the frequency gets shifted downwards, although
not dramatically. However, as already mentioned, the calculation is
limited to temperatures above $\sim 0.06 T_F$ (the critical
temperature $T_C$), while it is possible that the temperature in the
experiment was lower (the presence of a small superfluid region in the
center of the trap would not contradict the observation of the
collisionless frequency \cite{Urban2008}). Extrapolating the damping
curve obtained with the in-medium cross section to lower temperatures,
it seems that the result obtained with in-medium cross section is not
necessarily inconsistent with the experiment. Additional experimental
data points at higher (and known) temperatures could help to settle
this question.
%
\subsection{Scissors mode at $1/k_Fa = -0.45$}
\label{sec:scifin}
\begin{figure*}
\includegraphics[width=5.8cm]{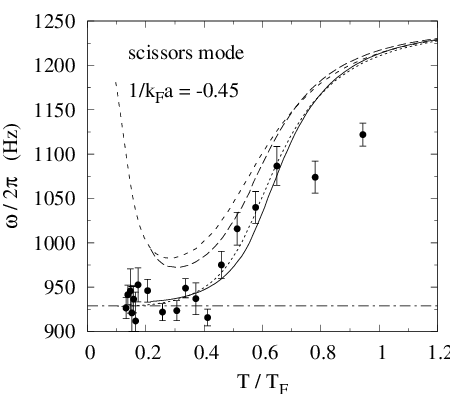}
\includegraphics[width=5.8cm]{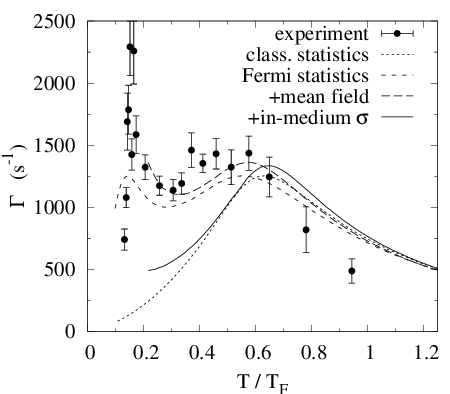}
\includegraphics[width=5.8cm]{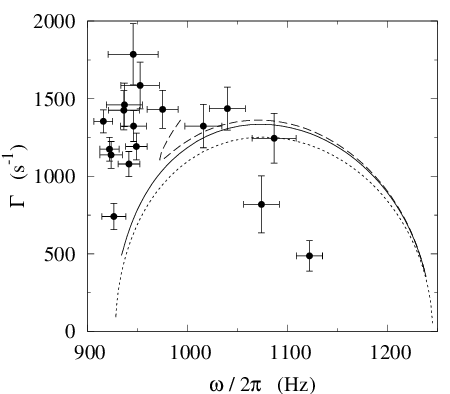}
\caption{\label{fig:scifin} Same as \Fig{fig:quafin}, but for the
 scissors mode and $1/k_F a = -0.45$. The data are from
 \Ref{Wright}. The dash-dotted line represents the hydrodynamic
 frequency, $\omega_{S,hd}=\sqrt{\omega_x^2+\omega_y^2}$}
\end{figure*}

%
Shortly after the quadrupole mode, the Innsbruck group studied the
scissors mode at $1/k_Fa = -0.45$ and at unitarity ($1/k_F a = 0$)
\cite{Wright}. In this experiment, the trap had frequencies $\omega_x
= 2\pi\times 830$ Hz, $\omega_y = 2\pi\times 415$ Hz, and $\omega_z =
2\pi\times 22$ Hz, and contained $N=400000$ $^6$Li atoms. The
frequency $\omega$ and damping $\Gamma$ were measured for constant
interaction strength as functions of the temperature. The experimental
data for the case $1/k_Fa=-0.45$, taken from \Ref{Wright}, are shown
in \Fig{fig:scifin}
together with various theoretical results. As in \Fig{fig:quafin}, we
display, in addition to $\omega$ and $\Gamma$ as functions of the
temperature, $\Gamma$ as a function of $\omega$.

The meaning of the different curves is the same as in the case of the
quadrupole mode discussed in \Sec{sec:quafin}. The most curious
result, which has already been observed in \Ref{BruunSmithScissors}
for the case $1/k_Fa=0$, is that already with classical statistics
(dotted line) one can reproduce quite well the observed frequencies:
At low temperature, the frequency is the hydrodynamic one
($\sqrt{\omega_x^2+\omega_y^2} = 2\pi\times 928$ Hz), and with
increasing temperature it rises towards the collisionless frequency
($\omega_x+\omega_y = 2\pi\times 1245$ Hz). What is most surprising is
that the agreement is very good at temperatures well below the
degeneracy temperature $T_F$, where the classical approximation is not
justified at all, while it fails at higher temperatures. In fact, the
high-temperature behavior of the frequency is not reproduced by any of
the theoretical calculations, which might be due to the importance of
the anharmonicity of the trap at high temperatures. The agreement
between the measured frequencies and those of a classical gas is,
however, purely accidental, as one can see by looking at the
corresponding damping rates. The classical statistics leads to a
damping which is much too weak at low temperatures (dotted line).

In fact, as in the case of the quadrupole mode discussed in
\Sec{sec:quafin}, the lack of Pauli blocking results in a high
collision rate, leading to a perfectly hydrodynamic behavior. The
inclusion of Pauli blocking (short dashes) strongly reduces the
collision rate and therefore increases the damping at low
temperatures, resulting in a very good agreement with the measured
damping rates (except near the peak at $T/T_F \sim 0.15$, which is
probably due to the superfluid-normal phase transition). Note that,
since the interaction is much stronger now than in the case of the
quadrupole mode discussed above, the collisionless regime is not
reached, although at low temperature the frequency increases strongly
towards the collisionless one. This increase of the frequency is not
observed in the experiment, because at these temperatures the system
is already in the superfluid phase and therefore its frequency stays
close to the hydrodynamic one, even if the collision rate is low. This
effect cannot be described in the framework of the simple Boltzmann
equation which does not include superfluidity. But also at higher
temperatures, the agreement of the frequencies obtained with Fermi
statistics (short dashes) with the data is not as good as that
obtained with Boltzmann statistics (dotted line).

The inclusion of the mean field (long dashes) leads to a small
reduction of the frequency, while the damping is slightly enhanced,
improving the agreement with the data. Since our calculation is
limited to the normal phase, the curves are restricted to temperatures
above $\sim 0.2 T_F$ (the critical temperature $T_C$). The frequencies
are now well reproduced for temperatures above $\sim 0.3 T_F$, while
they are still slightly too high between $\sim 0.2$ and $\sim 0.3
T_F$. The damping is in excellent agreement with the data for all
temperatures above $T_C$.

Finally, the inclusion of the in-medium cross section (solid lines)
leads to a big disappointment: The agreement with the data, in
particular for the damping, is completely lost. The results are very
close to those of the classical gas, similar to the findings of
\Ref{BruunSmithShearViscosity} for the shear viscosity of the unitary
gas and of \Ref{RiedlBruun} for different collective modes at
unitarity. The reason is that the enhancement of the cross section
(cf. \Fig{fig:xsection}) cancels the effect of Pauli blocking.

Apparently the present theory has a fundamental problem. Maybe the
quasiparticle approximation made in \Sec{sec:selfenergy} is too crude
(although the QP density profile coincides very well with the NSR
one): There might be important corrections due to energy and momentum
dependence of the self-energy \cite{Danielewicz}. Even the validity of
the Boltzmann equation itself might be questioned: The $T$ matrix
approximation can result in a long correlation time, leading to
non-Markovian (memory) effects \cite{Kremp}.
%
\subsection{Collective modes in the unitary limit}
\begin{figure*}
\includegraphics[width=5.8cm]{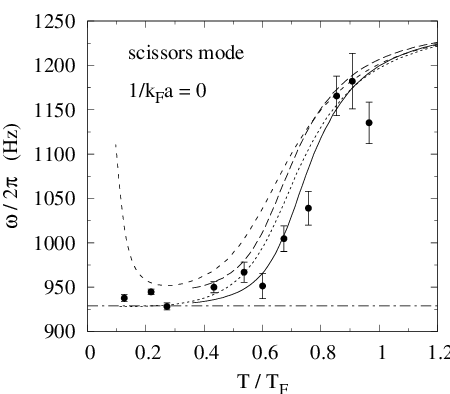}
\includegraphics[width=5.8cm]{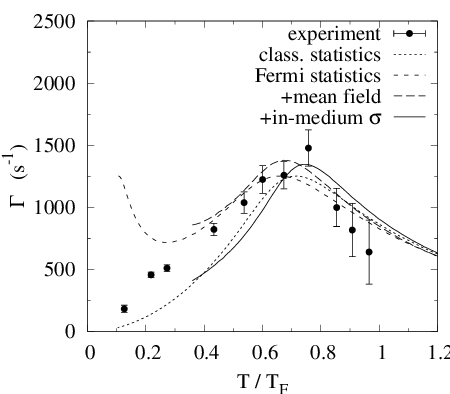}
\includegraphics[width=5.8cm]{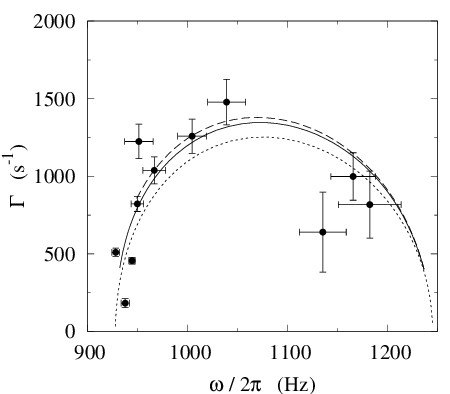}\\
\includegraphics[width=5.8cm]{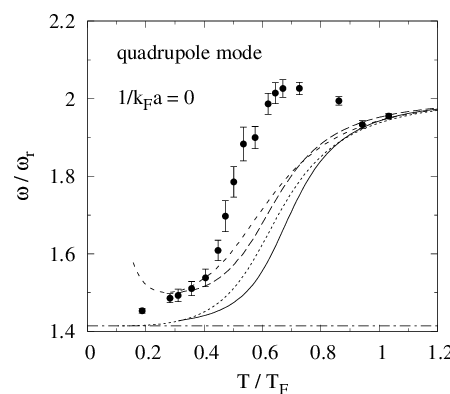}
\includegraphics[width=5.8cm]{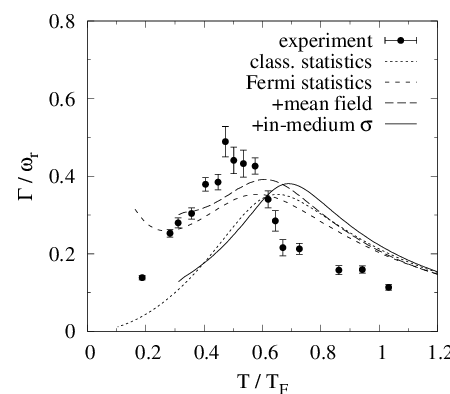}
\includegraphics[width=5.8cm]{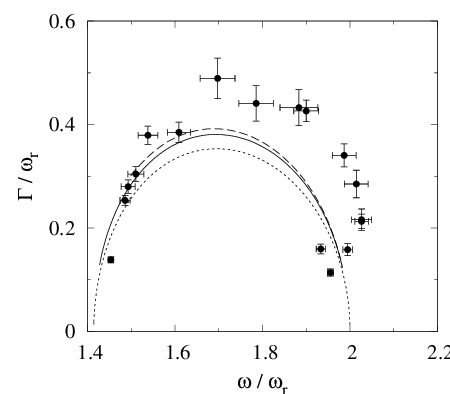}\\
\includegraphics[width=5.8cm]{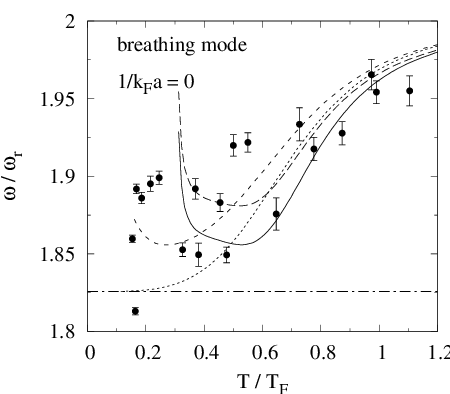}
\includegraphics[width=5.8cm]{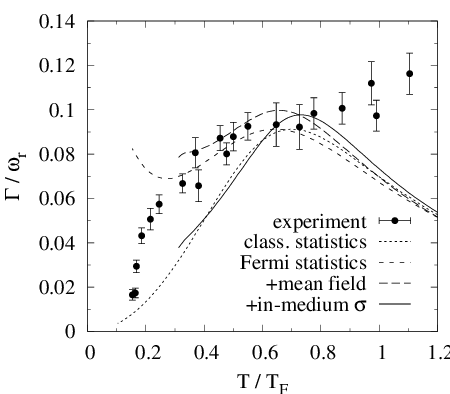}
\includegraphics[width=5.8cm]{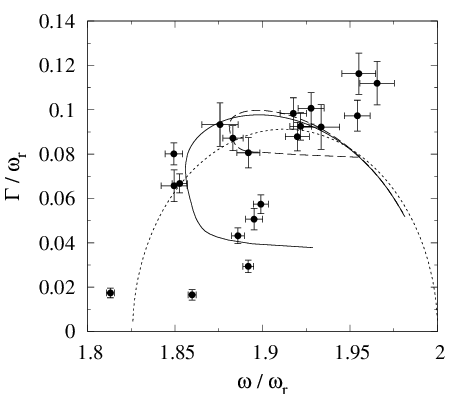}
\caption{\label{fig:uni} Same as \Fig{fig:quafin}, but for the
  scissors mode, the radial quadrupole mode, and the radial breathing
  mode (from top to bottom) in the unitary limit ($1/k_Fa = 0$). The
  experimental data are taken from \Ref{Wright} for the scissors mode
  and from \Ref{RiedlBruun} for the radial quadrupole and breathing
  modes. The dash-dotted lines represent the respective hydrodynamic
  frequencies of each mode.}
\end{figure*}

%
In this subsection we will finally show results for collective modes
in the unitary limit ($1/k_Fa = 0$). We will again compare with
experimental results obtained by the Innsbruck group, \Ref{Wright} for
the case of the scissors mode and \Ref{RiedlBruun} for the radial
quadrupole and breathing modes. In the experiment on the scissors
mode, the trap parameters were the same as those stated in the
beginning of \Sec{sec:scifin}. Our theoretical results and the
experimental data are shown in the first row of \Fig{fig:uni},
the representations are analogous to those in \Figs{fig:quafin} and
\ref{fig:scifin}.

The quadrupole mode was studied in an axially symmetric trap having
$\omega_r = 2\pi\times 1100$ Hz, $\omega_z = 2\pi\times 26$ Hz and
containing $N = 600000$ $^6$Li atoms \cite{RiedlBruun}. The
theoretical results and the experimental data are shown in the second
row of \Fig{fig:uni}. For the radial breathing mode, the trap
frequencies were somewhat higher, namely $\omega_r = 2\pi\times 1800$
Hz, $\omega_z = 2\pi\times 32$ Hz, but the number of atoms was again
$N = 600000$ \cite{RiedlBruun}. The corresponding theoretical results
and experimental data are displayed in the third row of
\Fig{fig:uni}.

In the case of the scissors mode (upper row of \Fig{fig:uni}), the
qualitative features of frequencies and damping as functions of
temperature are reasonably well reproduced by all approximations [even
  by the classical gas (dotted lines), as it happened already in
  \Sec{sec:scifin}]. As in the case of the scissors mode at $1/k_Fa =
-0.45$ discussed in \Sec{sec:scifin}, the data stay near the
hydrodynamic frequency at low temperature because of superfluidity,
which is not included in our theory. For temperatures above $T_C$,
which within our theory is given by $T_C \approx 0.3 T_F$, the
frequencies are best reproduced by the calculation including mean
field and in-medium cross section (solid line). Unfortunately, as it
happened already in the case $1/k_Fa = -0.45$, the strong enhancement
of the in-medium cross section leads to a damping which is much
smaller than the experimentally observed one. The damping as a
function of temperature is best reproduced by the calculation
including the mean field but using only the free cross-section in the
collision term (long dashes). However, if one looks at the plot
$\Gamma$ vs. $\omega$, both approximations result in curves which are
close to the data. This means that both approximations might be
compatible with the data, if it turned out that the temperature axis
was incorrect.

In this context it should be mentioned that the temperature
measurement in the experiment is not evident and in addition not
completely independent of the theoretical model used in the analysis.
For instance, in the analysis of \Ref{Wright}, the method introduced
by Thomas et al. in \Ref{ThomasKinast} was used, which requires, among
other things, the knowledge of a parameter $\beta$ determining the
``effective mass'' in the unitary limit. As a consequence, in addition
to the statistical error in the determination of the temperature,
there could be a sizable systematic error coming from theoretical
uncertainties. This observation underlines the usefulness of the
representation of $\Gamma$ vs. $\omega$, which is independent of the
temperature.

Let us now look at the radial quadrupole mode (second row of
\Fig{fig:uni}).  In this case, we have to admit that the differences
between the different approximations are much smaller than their
deviation from the experimental data, i.e., all approximations fail to
give a satisfactory description of the data. The mean field has only a
very small effect (difference between short and long dashes), and the
in-medium cross section (solid line) leads to an additional
deterioration as compared to the free one (long dashes). It seems as
if in the experiment the continuous transition from the hydrodynamic
to the collisionless case happened at a much lower temperature than in
any of the theoretical results. But also in the $\omega-\Gamma$ plot,
where the experimental data follow a very well defined curve, the
theoretical results are quite far from the data. Further studies
within an improved theoretical framework are needed.

Concerning the radial breathing mode (third row of \Fig{fig:uni}), it
is difficult to draw firm conclusions from the figures. As it was the
case for the scissors mode, the results obtained with mean field but
without in-medium cross section (long dashes) give a satisfactory
description of the damping between $T_C \approx 0.3 T_F$ and $\sim 0.8
T_F$, while the damping obtained with the in-medium cross section
(solid line) is too weak. Because of their strong scattering, the
frequency data seem to be compatible with any of the theoretical
curves, except for the sudden rise of the frequencies obtained with
mean field (long dashes and solid line) if one aproaches $T_C$ from
above. This rise of the theoretical frequencies does not come from a
reduced collision rate, leading to collisionless behavior, but from a
sudden increase of the interaction-dependent parameter $\chi^\prime$
(defined in Appendix \ref{app:breathing}). This effect is an artefact
of our QP approximation since it occurs when the density profile gets
flat at some point different from $\vek{r}=0$, which happens only with the
QP densities close to $T_C$, but not with the NSR densities.
%
\section{Conclusions}
\label{sec:conclusions}
%
In this paper, we studied the collective modes of a cold Fermi gas with
attractive interaction in the framework of the Boltzmann equation with
in-medium effects. Our starting point was the $T$ matrix approximation
(ladder resummation), which is known to give a satisfactory
description of the BCS-BEC crossover. We discussed the corresponding
single-particle self energy and in-medium scattering cross section
above the critical temperature $T_C$. Within the QP approximation,
which we have to make in view of the Boltzmann equation, the
self-energy acts effectively like an attractive mean field, but it
avoids the pathological problems of the Hartree
approximation. Although we used at the present stage the simplest
possible version of the QP approximation (neglecting energy and
momentum dependence of the self-energy), it reproduces well the
equilibrium density profiles obtained with the NSR formula, except
near unitarity close to $T_C$. The in-medium scattering cross section,
which is also obtained from the $T$ matrix, is strongly enhanced as
compared with the free one at low temperature. This is a precursor
effect of the transition to the superfluid phase.

We derived the frequencies and damping rates of different collective
modes (sloshing, scissors, radial quadrupole, and breathing modes)
with inclusion of the ``mean field'' and in-medium cross section. To
that end, we linearized the Boltzmann equation around equilibrium. A
first important result is that the frequency of the sloshing mode is
equal to the trap frequency even in the presence of the medium
effects, in accordance with the Kohn theorem. We stress that this
result can only be obtained if one uses in the linearized Boltzmann
equation the same mean field as in the calculation of the equilibrium
density profile.  [It therefore seems dangerous to calculate the
sloshing-mode frequency by inserting an interacting density profile
(e.g., a measured one), into an equation like Eq. (B2) of
\Ref{RiedlBruun} derived from the Boltzmann equation of a
non-interacting gas.]

The frequencies and damping rates of the different collective modes
were evaluated numerically as functions of the temperature and for
different values of the scattering length (parameter $1/k_F a$).
Because of the experimental uncertainty concerning the temperature, we
discussed in addition the ``hydrodynamic circles'', i.e., the damping
as a function of the frequency. It was shown that using classical
statistics one clearly cannot reproduce the observed collisionless
behavior of the modes at small temperatures. By the inclusion of Fermi
statistics (Pauli blocking), this problem is solved. Including the
mean field on top of Fermi statistics, which is the main topic of this
paper, has only a relatively small effect, but nevertheless it leads
to a significant improvement of the measured frequencies and damping
rates. For instance, we can for the first time quantitatively explain
the observed upwards shift of the quadrupole mode in the collisionless
normal phase at weak coupling ($1/k_Fa = -1.34$), which has attracted
a lot of attention. However, in the strongly interacting cases, in
particular in the unitary limit, the frequencies depend much more
strongly on the collision rate than on the mean field and the
mean field effects are of minor importance.

On top of that, we included the in-medium effects in the cross section,
which determines the collision rate (relaxation time). Since the
in-medium cross section becomes very large when one approaches the
critical temperature, this compensates the effect of Pauli blocking
and reduces the result of the full calculation to something comparable
with the result obtained for a classical gas, as already noted in
\Refs{RiedlBruun,BruunSmithShearViscosity}. Clearly, this is an up to now
unsolved problem and needs further examination. A first possible
extension of the present theory is to take into account the energy and
momentum dependence of the self-energy in a more involved QP
approximation \cite{Danielewicz}.

In addition, the method of taking moments of the Boltzmann equation,
although it proved very successful in the past, does not correspond to
a full solution of the Boltzmann equation. In particular, this method
is insufficient for a description of the damping due to the
anharmonicity of the potential (trap and mean field). In order to
improve the calculation in this respect, a more involved numerical
method for solving the Boltzmann equation is necessary. A very common
method is that using so-called ``test-particles''; in the context of
trapped fermionic atoms it was used, e.g., in
\Refs{Toschi2003, Toschi2004}.

Finally, a shortcoming of our present calculation is its limitation to
temperatures above $T_C$, whereas the most interesting experimental
results (e.g., extremely strong damping at a certain temperature) are
probably related to the transition to the superfluid phase. At least
in the weakly coupled regime, it should be possible to include the
effects of superfluidity following the approach of
\Refs{UrbanSchuck,Urban2007}. Work in these directions is in progress.
\begin{acknowledgments}
We are grateful to E.R. S{\'a}nchez Guajardo for providing us the data
of \Ref{Wright} with the corresponding temperature ratios $T/T_F$. One
of the authors (S.C.) acknowledges financial support provided by
Fondazione ``A.Della Riccia''.
\end{acknowledgments}
\appendix
\section{Virial theorem}
\label{app:virial}
%
Let us derive a relationship between equilibrium quantities which is
very useful for reducing the number of interaction dependent
parameters in the explicit expressions for the collective-mode
frequencies.

From the Boltzmann equation in equilibrium one obtains the following
property of the equilibrium distribution function:
\begin{equation}
\Big(\frac{\vek{p}}{m}\cdot \vek{\nabla}_r
  -\vek{\nabla}_r (V_T+U_\eq)\cdot\vek{\nabla}_p\Big)
  f_\eq(1-A f_\eq)=0\,.
\label{eq:proeq}
\end{equation}
Notice that this property was already used in order to derive
\Eq{eq:BoltzLin}.

In \Ref{MassignanBruun}, where the case $U=g\rho$ is considered for
the mean field, it is shown that multiplying equation (\ref{eq:proeq})
by $xp_x^2p_y^2$ and integrating over $\vek{r}$ and $\vek{p}$, one
obtains the virial theorem $E_\kin-E_\pot+3E_\inter/2=0$
\cite{Dalfovo,VichiStringari}, with
\begin{gather}
E_\kin = 2\int d^3r\, d^3p\, \frac{p^2}{2m}\,f_\eq\,,\\ 
E_\pot = 2\int d^3r\,V_T\rho_\eq\,,\quad \mbox{and}\\
E_\inter = g\int d^3r\rho_\eq^2\,.
\end{gather}
Here we want to show that the same can be done for any function
$U_\eq(\vek{r}) = U[\mu(\vek{r}),T]$ in order to obtain a generalized
virial theorem.

As mentioned in the end of \Sec{sec:LDA}, we define the rescaled
coordinates $\tilde{r_i}=r_i\omega_i/\bar{\omega}$, in terms of which
the trap potential reduces to $V_T=m\bar{\omega}^2\tilde{r}^2/2$, and,
consequently, also the density $\rho_\eq$ and the mean field $U_\eq$
become spherically symmetric in these coordinates.

It is then found that the generalized virial theorem is
\begin{equation}
E_\kin-E_\pot-\int d^3\tilde{r}\,
\rho_\eq\tilde{r}\,\frac{\p U_\eq}{\p\tilde{r}}=0\,.
\label{generalizedvirial}
\end{equation}
In the case $U_\eq=g\rho_\eq$ (Hartree approximation), the last term
can be integrated by parts, and the well-known result for the virial
theorem is recovered.

In the general case, let us define the parameter $\chi$ characterizing
the strength of the interaction as
\begin{equation}
\chi = -\frac{1}{E_\pot}\int d^3\tilde{r}\, \rho_\eq\tilde{r}\,\frac{\p
U_\eq}{\p\tilde{r}}\,.
\label{eq:chi}
\end{equation}
Then the virial theorem can be written as
\begin{equation}
\frac{E_\kin}{E_\pot}=1-\chi\,.
\label{eq:genVir}
\end{equation}
%
\section{Computation of the relaxation time}
\label{app:relaxation}
%
The equation determining the frequencies of the scissors mode contains
the parameter $\tau$, defined in \Eq{eq:tausci}). Its evaluation is
quite involved and we follow closely \Ref{Vichi} in order to reduce
the number of integrals. The intregral $I_S$ entering the definition
of $\tau$ can be most conveniently computed if one observes that
\begin{equation}
I_S = \frac{1}{10}\sum_{ij} I_{ijij} \,.
\end{equation}
The explicit expression for $I_S$ reads now
\begin{multline}
I_S = \frac{1}{10} \int d^3r\,d^3p\, d^3p_1\,
  d\Omega\,\frac{d\sigma}{d\Omega} \frac{|\vek{p}-\vek{p}_1|}{m}\\
  \times f_{\eq} f_{\eq 1}(1-Af_{\eq}^\prime)(1-Af_{\eq 1}^\prime)\\
  \times [p^4+(\vek{p}\cdot\vek{p}_1)^2-(\vek{p}\cdot\vek{p}^\prime)^2
  -(\vek{p}\cdot\vek{p}_1^\prime)^2]\,.
\label{ISthreelines}
\end{multline}
In order to reduce the number of integrals, one first introduces the
variables $\vek{k}=\vek{p}+\vek{p}_1$, $\vek{q} =
(\vek{p}-\vek{p}_1)/2$, and $\vek{q}^\prime =
(\vek{p}^\prime-\vek{p})/2$ (remember that $|\vek{q}| =
|\vek{q^\prime}|$). In terms of these variables, the factor in the
second line of \Eq{ISthreelines} becomes
\begin{multline}
\frac{1}{4A^2}\,\frac{1}{\cosh \beta(E-\mu_0)+
  \cosh \beta\vek{k}\cdot\vek{q}/2m}\\
\times \frac{1}{\cosh \beta(E-\mu_0)+
  \cosh \beta\vek{k}\cdot\vek{q}^\prime/2m}\,,
\end{multline}
with $E=k^2/4m+q^2/2m+V_T+U$. The factor in the third line of
\Eq{ISthreelines} reduces to $2q^4 - 2(\vek{q}\cdot\vek{q}^\prime)^2 +
(\vek{k}\cdot\vek{q})^2/2 - (\vek{k}\cdot\vek{q}^\prime)^2/2$. Note
that the last two terms do not contribute to the integral since they
are antisymmetric with respect to the interchange
$\vek{q}\leftrightarrow\vek{q}^\prime$. Let us now denote by
$\theta,\phi$ and $\theta^\prime, \phi^\prime$ the zenith and azimuth
angles of $\vek{q}$ and $\vek{q}^\prime$, respectively, with respect
to $\vek{k}$. The integrals over $\phi$ and $\phi^\prime$, which
appear only in the third line of \Eq{ISthreelines}, can be done
analytically and, writing $\gamma = \cos\theta$ and $\gamma^\prime =
\cos\theta^\prime$, we are finally left with
\begin{multline}
I_S = \frac{1}{20\pi^2 m}\int_0^\infty d\tilde{r}\, \tilde{r}^2
  \int_0^\infty dk\, k^2 \int_0^\infty dq\, q^7 \frac{d\sigma}{d\Omega}\\
  \times \int_{-1}^1 d\gamma\int_{-1}^1 d\gamma^\prime
  (1+2\gamma^2-3\gamma^2 \gamma^{\prime\,2})\\
  \times \frac{1}{\cosh \beta(E-\mu_0)+
  \cosh \beta kq\gamma/2m}\\
\times \frac{1}{\cosh \beta(E-\mu_0)+
  \cosh \beta kq\gamma^\prime/2m}\,.
\end{multline}
This five-dimensional integral is evaluated numerically using a
Monte-Carlo algorithm.
%
\section{Frequencies of the scissors mode}
\label{app:scissors}
%
In order to determine the frequency of the collective modes, one has
to take moments of the Boltzmann equation. It is thus useful to
rewrite \Eq{eq:BoltzLin}) as
\begin{multline}
f_\eq(1-Af_\eq)\dot{\Phi}-(\vek{\nabla}_p
  f_\eq) \cdot \vek{\nabla}_r \Big(T \Phi +
  \frac{dU_\eq}{d\rho_\eq}\delta\rho\Big) \\
  + T (\vek{\nabla}_r f_{\eq})\cdot\vek{\nabla}_p \Phi
 = -I[\Phi]\,,
\label{eq:BoltzLin2}
\end{multline}
and to denote the three terms on the lhs by $(i)$, $(ii)$, and
$(iii)$. When integrating over $\vek{r}$ and $\vek{p}$, the following
identities are useful:
\begin{gather}
\vek{\nabla}_p f_\eq = -\frac{\beta}{m}f_\eq(1-Af_\eq)\vek{p}\,,\\
\frac{\p f_\eq}{\p\mu} = \beta f_\eq(1-Af_\eq)
\Big(1-\frac{\p U_\eq}{\p\mu}\Big|_T\Big)\,,\\
\frac{\p}{\p\mu}\cdots\Big|_T = -\frac{1}{m\bar{\omega}^2\tilde{r}}
\frac{\p}{\p\tilde{r}}\cdots\,.
\end{gather}

As an example, we report in more detail the derivation of the modified
frequencies of the scissors mode. The corresponding trial function
$\Phi$ is given in \Tab{tab:Phi}.

The contributions of the term $(i)$ to the moments of Boltzmann equation are:
\begin{gather}
\int d^3r\, d^3p\, (i)\, xy =
  -\frac{i\omega E_\pot(1+\varphi_1)}{3\beta m^2\omega_x^2\omega_y^2}
  e^{-i\omega t} c_1\,,\\
\int d^3r\, d^3p\, (i)\, xp_y =
  -\frac{i\omega E_\pot}{3\beta\omega_x^2}e^{-i\omega t} c_2\,,\\
\int d^3r\, d^3p\, (i)\, yp_x =
  -\frac{i\omega E_\pot}{3\beta\omega_y^2}e^{-i\omega t}c_3\,,\\
\int d^3 r d^3p\, (i)\, p_xp_y =
  -\frac{i\omega m^2E_\kin}{3\beta}e^{-i\omega t}c_4\,,
\end{gather}
where the parameter $\varphi_1$ is defined as
\begin{equation}
\varphi_1 = \frac{1}{5E_\pot}\int d^3\tilde{r}\,\tilde{r}^2
\frac{\p\rho_\eq}{\p\tilde{r}}\frac{\p U_\eq}{\p\tilde{r}}
\frac{1}{1+\frac{1}{m\bar{\omega}^2\tilde{r}}\frac{\p U_\eq}{\p\tilde{r}}}\,.
\end{equation}

The density variation is given by
\begin{equation}
\delta\rho = e^{-i\omega t}c_1 xy\frac{1}{\beta}
\frac{\p \rho_\eq}{\p \mu}\frac{1}{1-\frac{\p U_\eq}{\p\mu}}\,.
\end{equation}
Using this, we can write for the contributions of $(ii)$ to the
moments of the Boltzmann equation:
\begin{gather}
\int d^3r\, d^3p\, (ii)\, xy = 0\,, \\
\int d^3r\, d^3p\, (ii)\, xp_y =
  \frac{(1+\varphi_1) E_\pot}{3 \beta m \omega_x^2} e^{-i\omega t}c_1\,, \\
\int d^3r\, d^3p\, (ii)\, yp_x =
  \frac{(1+\varphi_1)E_\pot}{3 \beta m \omega_y^2}e^{-i\omega t}c_1\,, \\
\int d^3 r\, d^3p\, (ii)\, p_xp_y = \frac{mE_\kin}{\beta}
  e^{-i\omega t}(c_2+c_3)\,.
\end{gather}

Finally, the $(iii)$ contributions to the moments 
of Boltzmann equation are:
\begin{gather}
\int d^3r\, d^3p\, (iii)\, xy =
  -\frac{E_\pot}{3\beta m} e^{-i\omega t}
  \Big(\frac{c_3}{\omega_y^2}+\frac{c_2}{\omega_x^2}\Big)\,,\\
\int d^3r\, d^3p\, (iii)\, xp_y =
  -\frac{m E_\kin}{3 \beta}e^{-i\omega t}c_4\,, \\
\int d^3r\, d^3p\ (iii)\, yp_x = -\frac{m E_\kin}{3 \beta}
  e^{-i\omega t} c_4\,, \\
\int d^3r\, d^3p\, (iii)\, p_xp_y = 0\,.
\end{gather}
With these results, the equations for the coefficients $c_i$ read
\begin{gather}
i\omega(1+\varphi_1) c_1 +m \omega_y^2c_2+ m \omega_x^2 c_3=0\,, \nonumber\\
(1+\varphi_1)c_1-im\omega c_2-(1-\chi)m^2\omega_x^2c_4=0\,, \nonumber\\
(1+\varphi_1)c_1-im\omega c_3-(1-\chi)m^2\omega_y^2c_4=0\,, \nonumber\\
c_2+c_3+m\Big(\frac{1}{\tau}-i\omega\Big) c_4=0\,. 
\end{gather}
The system has a solution if
\begin{multline}
\frac{i\omega}{\tau}[\omega^2-(\omega_x^2+\omega_y^2)]
  +\omega^4-2\omega^2(\omega_x^2+\omega_y^2)(1-\chi/2)\\
  +(\omega_x^2-\omega_y^2)^2(1-\chi)=0\,,
\end{multline}
which is \Eq{SciFrDa} with $\omega_{S,\hd}$ and $\omega_{S,\cl}$ given
by \Eqs{eq:whSciMF} and (\ref{eq:wcSciMF}).

\section{Frequencies of the breathing modes}
\label{app:breathing}

We consider a trap with frequencies $\omega_x=\omega_y=\omega_r$ and
$\omega_z=\lambda \omega_r$.

\begin{table*}
\label{tab:Breathing}
\begin{ruledtabular}
\begin{tabular}{c|cccccc}
& $x^2+y^2$ & $z^2$ & $xp_x+yp_y$ & $zp_z$ &  $p_x^2+p_y^2$ & $p_z^2$ \\
\hline
&&&&&&\\
$x^2+y^2$ & $\frac{2 i\omega (1+\varphi_1)}{m^2 \omega_r^2}$ &
$\frac{i\omega(1+\varphi_1)}{2m^2\omega_r^2}$ & $\frac{1}{m}$ & $0$ &
$i\omega$  & $\frac{i\omega}{2}$ \\
&&&&&&\\
$z^2$ &  $\frac{2 i\omega(1+\varphi_1)}{m^2 \omega_r^2}$ & 
 $\frac{3 i\omega (1+\varphi_1)}{m^2 \omega_r^2}$ & $0$ & $\frac{2}{m}$&
$2i\omega$ & $i\omega$\\
&&&&&&\\
$xp_x+yp_y$ & $\frac{2(1+2\varphi_1-\varphi_3)}{m}$ 
& $\frac{\varphi_1-\varphi_3}{m}$ & $-i\omega$ & $0$ &
$-2m\omega_r^2(1-2\chi+2\chi^\prime)$ & $m\omega_r^2(\chi-2\chi^\prime)$\\
&&&&&&\\
$zp_z$ & $\frac{2(\varphi_1-\varphi_3)}{m}$ & 
$\frac{2+3\varphi_1-\varphi_3}{m}$ & $0$ & $\frac{-i\omega}{\lambda^2}$
&$2m\omega_r^2(\chi-2\chi^\prime)$
&$-2m\omega_r^2(1-\frac{3}{2}\chi+\chi^\prime)$\\
&&&&&&\\
$p_x^2+p_y^2$ & $\frac{i\omega}{m^2\omega_r^2(1-\chi)}$ 
& $\frac{i\omega}{2m^2\omega_r^2(1-\chi)}$ 
& $-\frac{1}{m}$ & $0$ & $2i\omega -\frac{1}{3\tau}$ 
& $\frac{i\omega}{2}+\frac{1}{3\tau}$\\
&&&&&&\\
$p_z^2$ & 
 $\frac{i\omega}{m^2\omega_r^2(1-\chi)}$ &
  $\frac{i\omega}{2m^2\omega_r^2(1-\chi)}$ 
& $0$ & $-\frac{1}{m}$ & $i\omega+\frac{2}{3\tau}$ & 
$\frac{3}{2}i\omega -\frac{2}{3\tau}$
\end{tabular}
\end{ruledtabular}
\end{table*}
In \Tab{tab:Breathing}, each line is obtained by taking one moment of
the Boltzmann equation. For example, the third entry in the first
column ($1/m$) is the coefficient in front of $c_3$ if
\Eq{eq:BoltzLin2}, with $\Phi$ as given in the last line of
\Tab{tab:Phi}, is multiplied by $x^2+y^2$ and integrated over
$\vek{r}$ and $\vek{p}$. Notice that the coefficient $\lambda$ appears
only trough the $zp_z$ term. The new interaction dependent parameters
$\chi^\prime$ and $\varphi_3$ that enter in the table are defined as:
\begin{gather}
\chi^\prime = \frac{3}{2E_\pot}\int d^3\tilde{r}\, \rho_\eq^2 \frac{\p
U_\eq}{\p\rho_\eq}\,,\label{eq:chiPrime}\\
 \varphi_3 = -\frac{1}{E_\pot} \int d^3\tilde{r}\,
  \tilde{r}\rho_\eq \frac{\p U_\eq}{\p \tilde{r}}
  \frac{1}{1+\frac{1}{m\bar{\omega}^2\tilde{r}}\frac{\p U_\eq}{\p\tilde{r}}}\,.
\end{gather}
where $\chi^\prime$ reduces to $3 E_{int}/2 E_{pot}$ in the Hartree case.
From the determinant of the $6\times 6$ matrix given in
\Tab{tab:Breathing}, we obtain the equation for the frequencies,
\Eq{BreFrDa}. The frequencies in the collisionless limit are given by
\begin{equation}
\omega_{B,\cl\pm}^2=\omega_r^2\frac{a\pm\sqrt{a^2+b}}
  {16+25\varphi_1-25\chi(1+\varphi_1)}
\end{equation}
with 
\begin{align}
a = &25\chi^2(1+\lambda^2)(1+\varphi_1)+2[\chi^\prime(2+\lambda^2)
    (8+5\varphi_1) \nonumber\\
    &+(1+\lambda^2)(16+25\varphi_1)-4(2+\lambda^2)\varphi_3] \nonumber\\
    &+\chi[-10\chi^\prime(2+\lambda^2)(1+\varphi_1)
     -3(1+\lambda^2)(22+25\varphi_1) \nonumber\\
    &+8(2+\lambda^2)\varphi_3]\,,\\
b = &-4\lambda^2(2-\chi)[16+25\varphi_1-25\chi(1+\varphi_1)] \nonumber\\
    &\times[32+50\varphi_1+25\chi^2(1+\varphi_1)+6\chi^\prime(8+5\varphi_1)
     \nonumber\\
    &-3\chi(22+25\varphi_1+10\chi^\prime(1+\varphi_1)-8\varphi_3)-24\varphi_3]
    \,,
\end{align}
The hydrodynamic frequencies are
\begin{equation}
\omega_{B,\hd\pm}^2 = \omega_r^2 \frac{c\pm\sqrt{c^2+d}}
{3[16+25\varphi_1-25\chi(1+\varphi_1)]}\,,
\end{equation}
with 
\begin{align}
c = &25\chi^2(2+\lambda^2)(1+\varphi_1)
     +6\chi^\prime(2+\lambda^2)(8+5\varphi_1) \nonumber\\
    &+(5+4\lambda^2)(16+25\varphi_1)-24(2+\lambda)^2\varphi_3 \nonumber\\
    &+\chi[-157-175\varphi_1-30\chi^\prime(2+\lambda^2)(1+\varphi_1)
     +48\varphi_3 \nonumber\\
    &+\lambda^2(-116-125\varphi_1+24\varphi_3)]\,,\\
d = &-36\lambda^2[16+25\varphi_1-25\chi (1+\varphi_1)] \nonumber\\
    &\times[32+50\varphi_1+25\chi^2(1+\varphi_1)+6\chi^\prime(8+5\varphi_1)
     \nonumber\\
    &-3\chi(22+25\varphi_1+10\chi^\prime(1+\varphi_1)-8\varphi_3)-24\varphi_3]
     \,,
\end{align}
Note that since $b$ and $d$ are proportional to $\lambda$, the
low-lying limiting frequencies ($\omega_{B,\cl-}^2$ and
$\omega_{B,\hd-}^2$), corresponding to the axial breathing mode, tend to
zero in the limit of a very elongated trap ($\lambda\to 0$).

In absence of the mean field, $\chi=\chi^\prime=\varphi_1=\varphi_3=0$
and the frequencies reduce to the known expressions
\cite{MenottiPedri, MassignanBruun}
\begin{gather}
\omega_{B,\cl\pm}^2 =
  2\omega_r^2(1+\lambda^2\pm\sqrt{1-2\lambda^2+\lambda^4})\,,\\
\omega_{B,\hd\pm}^2 =
  \frac{\omega_r^2}{3}(5+4\lambda^2\pm\sqrt{25+16\lambda^4-32\lambda^2})\,.
\end{gather}


\end{document}